\documentclass[oldversion]{aa}  

\usepackage{graphicx}
\usepackage{natbib}

\begin{document}
 
\title{Dark Matter in Galaxy Clusters:\\ a Parametric Strong Lensing Approach}
   \titlerunning{}
   \authorrunning{Limousin et~al.}
   \author{Marceau Limousin\inst{1}, Benjamin Beauchesne\inst{2,3} \& Eric Jullo\inst{1}
      \thanks{Based on observations obtained with the \emph{Hubble Space Telescope}}
       }
   \offprints{marceau.limousin@lam.fr}

   \institute{
Aix Marseille Univ, CNRS, CNES, LAM, Marseille, France\\
Institute of Physics, Laboratory of Astrophysics, Ecole Polytechnique F\'ed\'erale de Lausanne (EPFL),
Observatoire de Sauverny, 1290 Versoix, Switzerland\\
ESO, Alonso de C\'ordova 3107, Vitacura, Santiago, Chile
              }

   %\date{Received...}
   
  \abstract
   {
We present a parametric strong lensing analysis of three massive galaxy clusters 
for which \emph{Hubble Space Telescope} imaging
is available, as well as spectroscopy of multiply imaged systems and galaxy cluster members.

Our aim is to probe the inner shape of dark matter haloes, in particular the existence of a
core. 

We adopt the following working hypothesis: any group/cluster scale dark matter clump introduced 
in the modelling should be associated with a luminous counterpart.
We also adopt some additional well motivated priors in the analysis, even if this degrades the
quality of the fit, quantified using the RMS between the observed and model generated images.

In particular, in order to alleviate the degeneracy between the smooth underlying component and the
galaxy scale perturbers, we use the results from spectroscopic campaigns by Bergamini et~al. (2019)
allowing to fix the mass of the galaxy scale component.

In the unimodal galaxy cluster AS\,1063, a cored mass model is favored
with respect to a non cored mass model, and this is also the case in
the multimodal cluster MACS\,J0416.
In the unimodal cluster MACS\,J1206, we fail to reproduce the strong lensing constraints using a 
parametric approach within the adopted working hypothesis. We then successfully add a mild perturbation 
in the form of a superposition of B-spline potentials which
allows to get a decent fit (RMS\,=\,0.5$\arcsec$), finally finding that a cored mass model is favored.
Overall, our analysis suggest evidence for \emph{cored} cluster scale dark matter haloes in these
three clusters.
These findings may be useful to interpret within alternative dark matter scenario, as self interacting
dark matter.

We propose a working hypothesis for parametric strong lensing modelling where the quest for the
best fit model will be balanced by the quest for presenting a physically motivated mass model,
in particular by imposing priors.
   }


   \maketitle
%________________________________________________________________

\section{Introduction}

Dark Matter (DM) is an elusive component which is supposed to largely dominate the mass budget
in astrophysical objects over a wide range of scales, in particular in galaxy clusters.
However, more than 80 years after the first \emph{indirect} evidence for DM in galaxy clusters (Zwicky 1937), we have no definitive
clues about its existence even though it is sometimes taken for granted.
Indeed, evidence for DM are \emph{indirect only} and no well understood and characterized particle detector has
detected it yet, despite intense works amongst the community 
\citep[see, \emph{e.g.},][]{Axions_review,direct_detection}.
As long as such a direct detection has not been reliably achieved, DM remains an hypothesis. 

In this paper, we \emph{assume} the DM hypothesis and we focus on its distribution on 
cluster scales, using \emph{parametric} strong lensing techniques.
Indeed, strong lensing (SL) is an essential probe of the DM distribution in the centre of
galaxy clusters, where the mass density is so high that space time is locally deformed such that
multiple images of background sources can form, providing valuable constraints on the mass
distribution.

Parametric SL mass modelling relies on the following working hypothesis, supported by N-body simulations:
a galaxy cluster is an object composed of different mass clumps, each component being associated
with a luminous counterpart and which can (to some extent) be described
parametrically. 
One advantage of parametric SL modelling is that the description of these mass clumps
can be directly compared with theoretical expectations.
However, a parametric description is sometimes not accurate nor adapted, emphasizing the
limit of parametric mass modelling and the need for more flexible approaches. 

We usually have two types of mass clumps: cluster scale DM clumps (whose typical projected mass within
a 50$\arcsec$ aperture is of order 10$^{14}$M$_{\sun}$ at $z\sim$\,0.2) and galaxy scale DM clumps associated with
individual galaxies. Added to that description of the dominant DM component, one can also consider in the
modelling mass components associated with the X-ray gas \citep{danka,Bonamigo_2018}.

Parametric SL modelling displays interesting and puzzling features.

Let us mention two of them that will be addressed in this paper.
First, it sometimes requires DM clumps whose position do not coincide with any
luminous counterpart. 
This is the case in complicated merging clusters, for example
MACS\,J0717 \citep{Limousin2016} or Abell~370 \citep{Lagattuta_2019},
where parametric mass
modelling might not be the best modelling method.
This also happens in apparently unimodal
clusters (\emph{e.g.} MACS\,J1206 studied in this work).
These "dark" clumps are usually added in order to improve significantly the fit
but the \emph{physical interpretation} of these clumps is not straightforward and, taken for
granted, their inclusion in the mass budget might be misleading: "are we \emph{really} 
witnessing a dark clump?".
These "dark" clumps are more likely to express the limitations of parametric mass modelling in some clusters. In this case, the advantage of parametric mass modelling mentioned above might lose their
relevance.

Second, large core radii (sometimes larger than 100\,kpc) are sometimes reported in parametric SL studies
\citep[\emph{e.g.}][]{sand04,locuss,Newman_2013,Lagattuta_2019,Richard_2021},
which have some implications for cosmological models
\citep[see discussion by][L16 hereafter.]{Limousin2016}.

Finally, some mass models do require non negligible external shear component ($\gamma_{\rm ext}$) in order to significantly
improve the goodness of the reconstruction, but the physical origin of this external shear is not always
clear,
for example in Abell~370 \citep[$\gamma_{\rm ext}$=0.13,][]{Lagattuta_2019}, in 
MACS\,J0329 and RX\,J1347 \citep[$\gamma_{\rm ext}$=0.07 and 0.10 respectively,][]{Caminha_2019},
in MACS\,J1206 \citep[$\gamma_{\rm ext}$=0.12,][]{Bergamini_2019}, in Abell~2744
\citep[$\gamma_{\rm ext}$=0.17,][]{Mahler_2017} and in SDSS\,1029 \citep[$\gamma_{\rm ext}$=0.09,][]{Acebron_2022}.
This component can have either a physical origin or might be compensating for the limitations of the parametric modelling.
In practice, the origin of this external shear can be missed since it can be due to substructures located far from
the cluster core \citep{Acebron_2017} in regions not covered in narrowly targeted observations as the one carried out with the HST.

Interestingly enough, some alternatives to the current cosmological $\Lambda$ Cold Dark Matter ($\Lambda$CDM) scenario allow for the formation of cores and offsets between luminous and dark 
components at galaxy cluster scale, as self interacting dark matter (SIDM) models.
For a recent review on the SIDM alternative, see \citet{SIDM_review}.

A thorough SIDM investigation of the size of the core radii in the galaxy cluster
mass regime is still lacking,
but we have some hints about its order of magnitude.
With low statistics (50 haloes resolved),
\citet{Rocha_2013} do quantify haloes core sizes in SIDM simulations with
a cross section equal to 1\,cm$^2$/g to be in the range 55-90 kpc.
Their simulations with a cross section equal to 0.1\,cm$^2$/g are not
resolved enough to measure a core radius, but they predict core radii to be
in the range 10-12 kpc.
\citet{Robertson_2017} report core radii smaller than 40\,kpc.
Such an upper bound is also found when self interaction is "frequent" instead of "rare"
\citep{Fischer_2021}.

Regarding offsets between stellar and DM components in the SIDM scenario, studies
have shown that such offsets should be small. In the case of the merging
Bullet Cluster where DM clumps positions are well constrained by SL, no such offsets has been
detected, providing an upper limit on the self interacting cross section of DM equal to
1.25\,cm$^2$/g (Randall 2008).

According to simulations by \citet{Kim_2017}, in merging clusters, a maximal galaxy
DM offset of 20\,kpc forms for a self interaction equal to 1\,cm$^2$/g.
They also find that the maximal one before completely disrupting
the haloes is $\sim$100\,kpc in an ideal case.
\citet{Robertson_2017}, presenting SIDM simulations with anisotropic scattering,
report DM-galaxy offsets smaller than 10\,kpc.
\citet{Fischer_2021}, in simulations of SIDM equal mass mergers
with frequent self interactions, claim that frequent self interactions can cause
much larger offsets than rare self interactions.
These offsets do vary with time and are non zero after the first pericentre passage.
For a self interaction of 1.0\,cm$^2$/g they
are found to be smaller than $\sim$\,100\,kpc (versus 50\,kpc for rare self interactions
of the same strength).
\citet{Harvey_2019} investigated the median BCG offset for different dark matter cross sections using the BAryons and HAloes of MAssive Systems simulations \citep[BAHAMAS,][]{McCarthy_2016} and found the median offset to be below 10 kpc.

To conclude, within an SIDM scenario, core radii should be smaller
than $\sim$\,100\,kpc. 
Besides, any
offset between stellar and DM components should be smaller than $\sim$\,100\,kpc.

The main aim of this paper is to investigate the shape of DM at the galaxy cluster scale, in
particular its central shape and the eventual presence of a core, following the work
presented by L16.

Constraints on DM properties, in particular its central shape, derived from SL 
have been hampered by the degeneracies inherent to this
technique.
Strong lensing being sensitive to the total projected mass, it can sometimes be difficult to get
insights into the underlying central DM distribution, suggesting the need for robust and well motivated
priors in order to break these degeneracies.

One of the main degeneracy is the one between the \emph{smooth} and the 
\emph{galaxy scale} DM components.
L16 have shown that, even in the Hubble Frontier Fields era where up to hundreds
of multiple images can be detected, this degeneracy still exists and prevents us from getting insights
on the DM distribution, in particular on its central shape and the presence of a core.

A promising avenue is to use priors on the galaxy scale perturbers.
Recent work by \citet[][B19 hereafter]{Bergamini_2019} propose to use kinematics of cluster members in order to alleviate this degeneracy by
providing well motivated priors on the mass of galaxy scale perturbers
that are used in the SL mass modelling.

In this paper, we revisit the mass models of three clusters for which 
B19 have derived reliable constraints on the galaxy scale
perturbers to be used as priors in the SL modelling.
Our main aim is to investigate whether including these priors can effectively
break the above mentioned degeneracy and help in providing constraints 
on the inner shape of the underlying DM distribution.
More precisely, following L16, we aim to see if we can disentangle between a cored and a 
non cored mass model.

Other priors will also be considered, which are aimed at presenting physically motivated 
mass models. In particular we \emph{impose light and mass to coincide within a few arcseconds.}
At the redshift of the clusters considered in this work, this translates into a few dozen of kpc,
\emph{i.e.} of the order of what is allowed by SIDM scenarios.

We will see that imposing priors sometimes leads to decreasing the quality of the SL fit.
This leads to the following relevant (open) question: \emph{"What is the balance between improving the fit and presenting a mass
model which is physically relevant and reliable?"}
This work also aims at providing some hints about this question.

\section{Methodology}

\subsection{dPIE profile}
Dark matter mass clumps are described using a dual Pseudo Isothermal
Elliptical Mass Distribution (dPIE profile).
We refer the reader to \citet{mypaperI} and \citet{ardis2218} for a description of
this mass profile. Here we give a brief overview.
It is parametrised by a fiducial velocity dispersion $\sigma$, a core radius $r_{\rm{core}}$ and a scale radius $r_s$, sometimes referred to as cut radius in other publications using this profile.
We prefer using $r_s$ instead of $r_{\rm cut}$. Indeed, it can be shown that
this scale radius corresponds to the radius containing half the 3D mass 
\citep{ardis2218}, hence $r_{\rm cut}$ can be misleading.
Between $r=0$ and $r=r_{\rm{core}}$, the mass density is constant. 
Then between $r=r_{\rm{core}}$ and $r=r_s$, the mass density is isothermal, then it falls as $r^{-4}$.
For the large scale DM clumps, we fix the scale radius to an arbitrary large value.

\subsection{Strong Lensing}
The mass models presented in this paper comprise large-scale DM haloes,
as well as perturbations associated with individual cluster galaxies, all being 
characterized using a dPIE profile.
For the modelling of individual cluster galaxies, we use the results of B19 and refer the reader to that paper
for a full description.

The optimisation is performed in the image plane using the \textsc{Lenstool} software (Jullo et~al. 2007).
The goodness of the fit is quantified using the RMS between the predicted and the observed multiple images.

The positional uncertainty of the images is an important ingredient for the $\chi^2$ computation.
It affects the derivation of errors, in the sense that smaller positional uncertainties can result
in smaller statistical uncertainties, which may be underestimated.
Depending on the mass model explored, we set the positional uncertainty to a value of the order of the image plane RMS,
in order to attain a reduced $\chi^2$ of order 1.
In practice, a positional uncertainty equal to 1$\arcsec$ is set for all clusters.

When investigating a mass model, the \textsc{rate} parameter that controls the speed of convergence in the MCMC sampler \citep{jullo07} is decreased until the results are stable
in the sense that they do not depend on the \textsc{rate} value, suggesting that the parameter space
is well enough sampled.

\subsection{Priors}

Priors in parametric SL mass modelling are important and worth including when they are
well motivated.
The choice of priors can definitely lead the best mass model one will infer from a SL
analysis and therefore have to be taken with care and with criticism.
We consider the following priors that apply to all models presented in this work.
Some additional priors are mentioned where relevant.

\paragraph{Large scale DM clumps.}
In this work, we force the positions of the large scale DM clumps to coincide with 
a luminous counterpart. When a clear central dominant galaxy is present (\emph{e.g.} in the centre
of AS\,1063), the position of the main DM clump is allowed to vary within the central light
distribution (typically a few arcseconds from the centre of the light distribution).
For mass clumps associated with a galaxy group 
(\emph{e.g.} the north-east galaxy group in AS\,1063), the position is allowed to vary
within the light distribution of the galaxy group or of the dominant galaxy (drawn by a 
square on each figure). This also corresponds to a few arcseconds ($\sim$ 20\,kpc at $z=0.4$). 
Note that this is the order of magnitude of the offsets allowed by SIDM models.

The ellipticity is forced to be smaller than 0.7, motivated by the results from
numerical simulations \citep{Despali_2016}, whereas it was allowed to reach 0.9 in the models by 
B19.

\paragraph{External shear.}
Any external shear taken into account in this work is forced to be smaller than 0.1, since there is not any
obvious massive neighbouring mass component that could account for such an external shear.

\paragraph{Galaxy Scale perturbers.}

We refer the reader to B19 for the discussion of the photometric and spectroscopic
data, in particular the observations used to measure the line-of-sight
velocity dispersion of cluster members which allow B19 to propose a prior on
the mass of galaxy scale perturbers, which we use here.
For MACS\,0416, we refer the reader to \citet{Bergamini_2021} which presents more data
than B19.

In short, B19 use VLT/MUSE integral-field spectroscopy in the cluster cores to measure the 
stellar velocity dispersion of 40-60 member galaxies per cluster. With these
data, they determine the normalization and slope of the galaxy Faber-Jackson relation 
between the luminosity and the velocity dispersion in each cluster.
Then they use these parameters as a prior for the scaling relations of the sub-halo 
population in the SL modelling. 

Moreover, the scale radius of an L$^{*}$ galaxy scale perturber is forced to be in the range
[5, 150\,kpc] (instead of [0, 250\,kpc] in B19), which is more in line
with galaxy-galaxy lensing results \citep{Monna_2016,mypaperII,priya4}.

\paragraph{X-ray gas.}
We take explicitly into account the gas component, using the results by \citet{Bonamigo_2018}.
In these three clusters, \citet{Bonamigo_2018} analyse the X-ray gas and describe its contribution
to the mass budget by a superposition of dPIE profiles.

\subsection{What is a "good" SL fit?}
SL studies usually quantify the goodness of their fit using the RMS between the
observed and predicted positions of the multiple images.
The lower this value, the better the SL fit.
The border between a "good" and a "bad" fit is somehow subjective.
Within the SL modelling community, we use to consider that an RMS lower than
0.5$\arcsec$ constitutes a good fit, and that an RMS lower than 1$\arcsec$ is still acceptable.
Indeed, considering recent parametric 
SL studies of massive clusters using space based data and spectroscopic 
confirmation of multiply imaged systems, RMS between 0.15$\arcsec$ and 1.0$\arcsec$ are reported
\citep[see, \emph{e.g.},][]{RELICS_2018,Cibirka_2018,Caminha_2019,Mahler_2019,Lagattuta_2019,Rescigno_2020,Acebron_2020,Richard_2021,Zitrin_2021}.

This consideration might also depend
on the complexity of the cluster: a multimodal complex cluster is usually more
difficult to model parametrically compared to a unimodal relaxed cluster.

Another related issue we face in this work aiming at comparing mass models is the following:
what is the minimum RMS between two models required in order to disentangle between
them?
Once again the answer to this question is subjective.
In MACS\,J0717 (L16), 
the RMS difference between a cored and a non cored mass model
was equal to 0.5$\arcsec$ (in favor of the cored model) and we argued that this difference was not large enough to
disentangle between both models.
Indeed it is comparable to or smaller than that due to an image mis-identification
\citep[see, \emph{e.g.} the case of image 3.3 in Abell~2744 discussed in][]{jauzac2744}, or to the
difference caused by an unaccounted for structure along the line of sight \citep{host}.

In this paper, we \emph{assume} that a good fit is attained when the RMS is lower than
1$\arcsec$, and that a difference of RMS of 2$\arcsec$ is large enough in order to
disentangle between two mass models with a good confidence.

\section{AS\,1063}

\begin{figure*}
\begin{center}
\includegraphics[scale=1.0,angle=0.0]{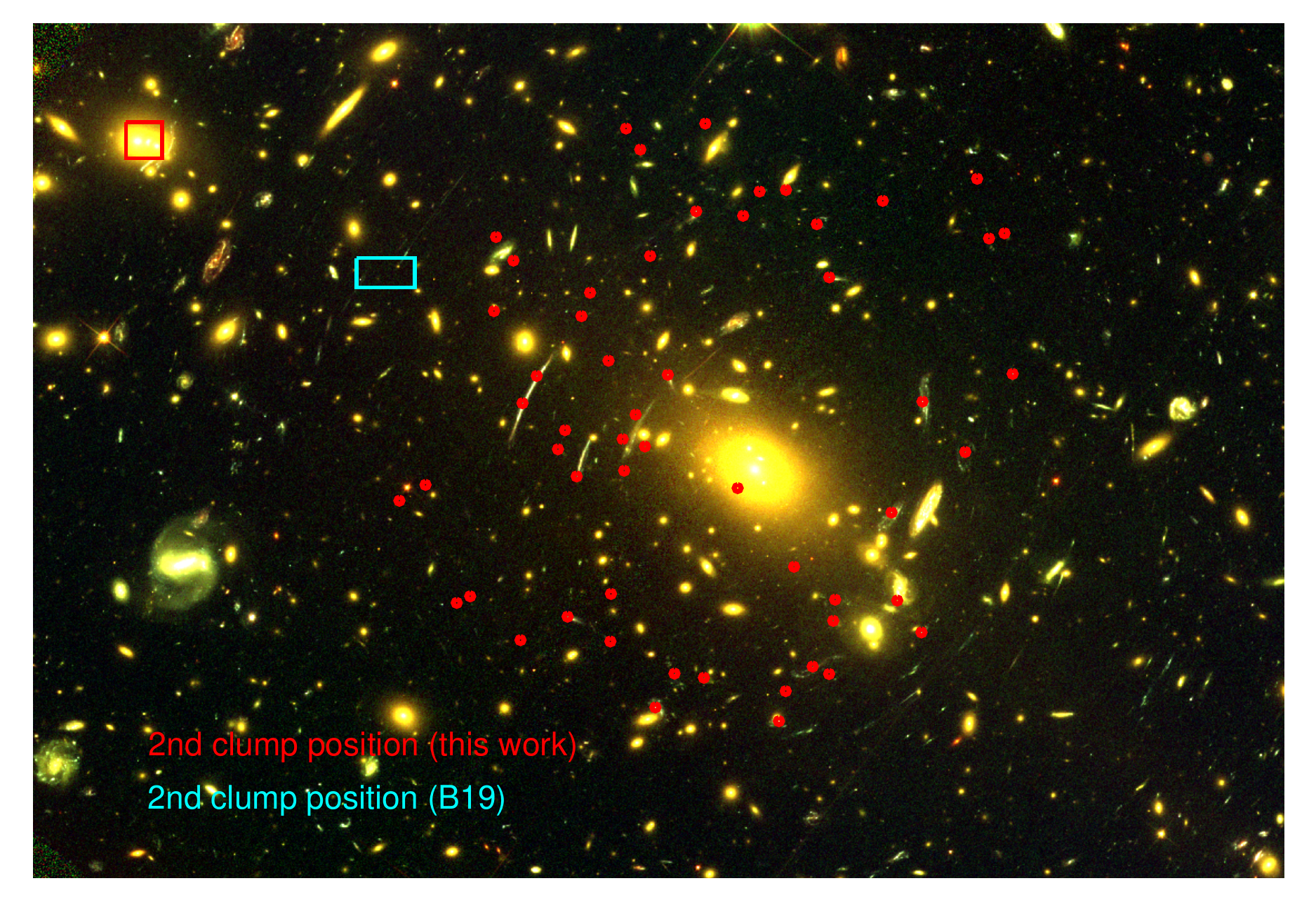}
\caption{Core of AS\,1063. Red circles represent the multiple images used in this work (see B19
for details). The boxes represent the location of the second mass clump associated with the north-east
group. Size of the field equals 150$\arcsec$\,$\times$\,96$\arcsec$. North is up, east is left.}
\label{1063field}
\end{center}
\end{figure*}

Abell AS\,1063 (a.k.a. RXJ 2248.7-4431), at redshift 0.348, has been observed deeply
as part of the CLASH \citep{clash} and Hubble Frontier Fields \citep[HFF,][]{Lotz_2017} programs.
It is the only HFF cluster which looks unimodal and which is clearly dominated by a bright central
galaxy.

If the dynamical state of the cluster has been debated, recent work combining X-ray and radio 
observations points out to a disturbed state, AS\,1063 being in an early stage of merging
\citep[see][and references therein]{Rahaman_2021}.

\subsection{Revisiting Bergamini et~al. (2019) model}

Following B19, we consider 55 multiple images coming from 20 sources (all spectroscopically
confirmed).

We start by reproducing the B19 mass model, which includes an elliptical dPIE
profile associated with the BCG,
and a circular dPIE whose position is left free to vary in an area of 
150$\arcsec$\,$\times$\,120$\arcsec$ centred in the north-east region of the 
cluster where we can observe a galaxy group which generates additional strong
lensing features for which no spectroscopic information is available, hence
not considered.
This mass clump has a vanishing core radius and a scale radius fixed 
to an arbitrary large value.

We do find the same parameters within the error bars and the same
RMS, equal to 0.55$\arcsec$.

This mass model is able to reproduce accurately the observational constraints. However the
location of the second clump does not coincide at all with 
the galaxy group in the North East (separation of 40$\arcsec$), although it was introduced 
to be associated with it (see cyan box on Fig.~\ref{1063field}).

We revisit the B19 mass model.
As mentioned above, we consider that having a mass clump whose position is not associated with a light 
concentration
is not satisfactory. We force the position of this second mass clump to 
coincide with the light distribution of the most luminous galaxy of this group (red square on Fig.~\ref{1063field}).
We get an RMS of 0.67$\arcsec$. 
If the RMS gets larger by 0.12$\arcsec$, we prefer this mass model 
since each mass clump is associated with a luminous counterpart.

We then add an external shear component.
This decreases the RMS to 0.53$\arcsec$.
The strength of this external shear equals 0.03 and its position angle 40 degrees, pointing
out to the north west direction.
If it lowers the RMS by 0.14$\arcsec$, we prefer not including it in the mass model used
to probe the underlying DM distribution.
The main reason is that we do not observe any clear interpretation to this component
(\emph{e.g.} a mass concentration in the north west), therefore including this external shear
might compensate for the limitations of the parametric modelling and eventually dilute the
conclusions of our study, \emph{i.e.} to disentangle between a cored and a non cored mass model.
We will test and discuss that further.

Finally, we try a model for which the underlying mass distribution is described by a single mass
clump, and find an RMS of 0.80$\arcsec$.
We prefer a two clumps mass model (without external shear) since it has a physical
interpretation.
It constitutes the reference model from which we continue our investigations.
Its parameters are given in Table~\ref{table}. We note that the core radius is equal to 89.5$\pm$5.5\,kpc, in agreement with
previous studies.

\subsection{Non cored mass model: a peaky DM distribution?}

Following L16, we do investigate here if a non cored mass model for the main DM clump
is also able to reproduce the observational constraints.
We redo the modelling imposing the core radius to be smaller than 10\,kpc, and obtain
an RMS equal to 3.83$\arcsec$.
Parameters are given in Table~\ref{table}.
The difference of RMS between the reference cored mass model and the non cored model
equal 3.16$\arcsec$, which we consider large enough to favor a cored mass model.
For this modeling, we have put a prior on the \emph{velocity
dispersion} of the second north east mass clump.
Indeed, in the first attempts of testing a non cored mass model, we found that the velocity dispersion of
the second mass clump was reaching values as high as 765\,km\,s$^{-1}$ (instead of
350\,km\,s$^{-1}$ for a cored mass model), which is somehow unrealistic
given the optical richness and the absence of an X-ray associated emission (Bonamigo et~al. 2018).
Therefore, we decided to limit the velocity dispersion of the second mass clump to 450\,km\,s$^{-1}$.
As shown in Table~\ref{table}, the value of this parameter is always stuck to this upper bound
of the prior in non cored models.
The RMS obtained without using this prior is 2.56$\arcsec$; this would lead to a difference of
RMS equal to 1.89$\arcsec$.
This highlights the importance of physically motivated priors.

\begin{table*}
\begin{center}
\begin{tabular}{ccccccccccc}
\hline \\*[-1mm]
Model & $\Delta$\,\textsc{ra} & $\Delta$\,\textsc{dec} &  $e$  & $\theta$ & $\sigma$  & $r_{\rm core}$ & L$^{*}$ galaxy $\sigma$ & L$^{*}$ galaxy $r_s$ & RMS & $\sigma$ (2nd clump)\\
      &      $\arcsec$        &    $\arcsec$           &       &          & km\,s$^{-1}$ &  kpc     &  km\,s$^{-1}$             &      kpc               & $\arcsec$ & km\,s$^{-1}$  \\ 
\hline \\*[-1mm]
Cored  &  1.4$\pm$0.5  & -0.9$\pm$0.4  & 0.64$\pm$0.01  & 141.1$\pm$0.5  &  1151$\pm$14  & 89.5$\pm$5.5   &  302$\pm$14  & 43$\pm$15.5 &  0.67$\arcsec$ & 350$\pm$50   \\   
\hline \\*[-1mm]
Non Cored  &  1.2$\pm$0.5  & 1.5$\pm$0.6  &  0.7$^{*}$  &  138.9$\pm$0.7  & 963$\pm$5 &  10$^{*}$  &   303$\pm15$ & 5$^{*}$ & 3.83$\arcsec$  & 450$^{*}$ \\   
\hline \\*[-1mm]
\smallskip
\end{tabular}
\end{center}
\caption{dPIE parameters inferred for the cored and the non cored models for the main clump describing AS\,1063 (except the last column which corresponds to the velocity dispersion of the
isothermal second clump), which has an arbitrary large scale radius $r_s$. 
Galaxy scale perturbers are described by dPIE profiles with vanishing core radii.
Coordinates are given in arcseconds relative to $\alpha$\,=\,342.18321, $\delta$\,=\,-44.530878; 
$e$ and $\theta$ are the ellipticity and position angle of the mass distribution.
Error bars correspond to the 1$\sigma$ confidence level.
Parameters with $*$ are stuck to a bound of the allowed prior.}
\label{table}
\end{table*}

\subsection{Cosmology}

In this work, we assume that the Universe is described by a flat Universe with $w_{\rm X}$\,=\,-1 and $\Omega_{\rm M}$\,=\,0.3.\\

We run a cored and a non cored mass model, letting $w_{\rm X}$ and $\Omega_{\rm M}$ free.
For each model, we compare the values of the optimized output cosmology with the values of the reference
cosmology mentioned above.
Our goal is to see if this test can help to discriminate between a cored and a non cored mass model.
For the cored model, we get an RMS of 0.62$\arcsec$ (instead of 0.67$\arcsec$ for the reference model) and the reference cosmology is retrieved.
For the non cored model, we get an RMS of 3.35$\arcsec$, and we do not retrieve the reference cosmology.
In particular, $\Omega_{\rm M}$ is stuck to the lower bound allowed, i.e. 0. 
This test gives further credit to the non cored mass model.
Note that we do not aim to constrain cosmology here but to provide an additional test in order 
to disentangle between a cored and a non cored mass model.

\subsection{External shear}
As mentioned above, an external shear component can compensate for the limitations 
of the parametric
modelling and dilute the discrimination between a cored and a 
non cored mass model.
The cored mass model with an external shear gives an RMS of 0.53$\arcsec$.
For the non cored mass model, allowing an external shear component lowers the RMS to
3.07$\arcsec$ (versus 3.75$\arcsec$ when no external shear is allowed).

With the inclusion of an external shear component, the RMS difference equals to 2.54$\arcsec$.
This is 0.5$\arcsec$ less discriminating compared to the case without any external shear.

\subsection{Degeneracies with the BCG}
We remove the BCG from the galaxy catalog and optimise it explicitly, in order to investigate
any degeneracies between the main DM clump core radius and the BCG parameters.
Its position is allowed to vary within $\pm$ 4$\arcsec$ from its centre and its core radius
between 1 and 50 kpc.
We use the values of B19 to put constraints on its velocity dispersion.
B19 provide a gaussian prior with a mean and a standard deviation. We consider the same
mean and the double of their standard deviation in order to allow for more freedom.
We get an RMS equal to 0.64$\arcsec$.
The optimized position is consistent with the centre of the light distribution.
The ellipticity is unconstrained.
Constraints on the core radius are very loose, the output PDF basically reproducing the prior, 
with a preference for the lower values. Therefore we see no degeneracies between the core radius of
the BCG and the core radius of the main DM clump.

The value of the main DM clump core radius is consistent with the reference model.

\section{MACS\,0416}

MACS\,0416, at redshift 0.396, has been observed deeply as part of the HFF program.

\subsection{Revisiting Bergamini et~al. (2019) model}

Recently, B19 and \citet{Bergamini_2021} (B21 hereafter) presented two mass models for this cluster.

\begin{figure*}
\begin{center}
\includegraphics[scale=1.0,angle=0.0]{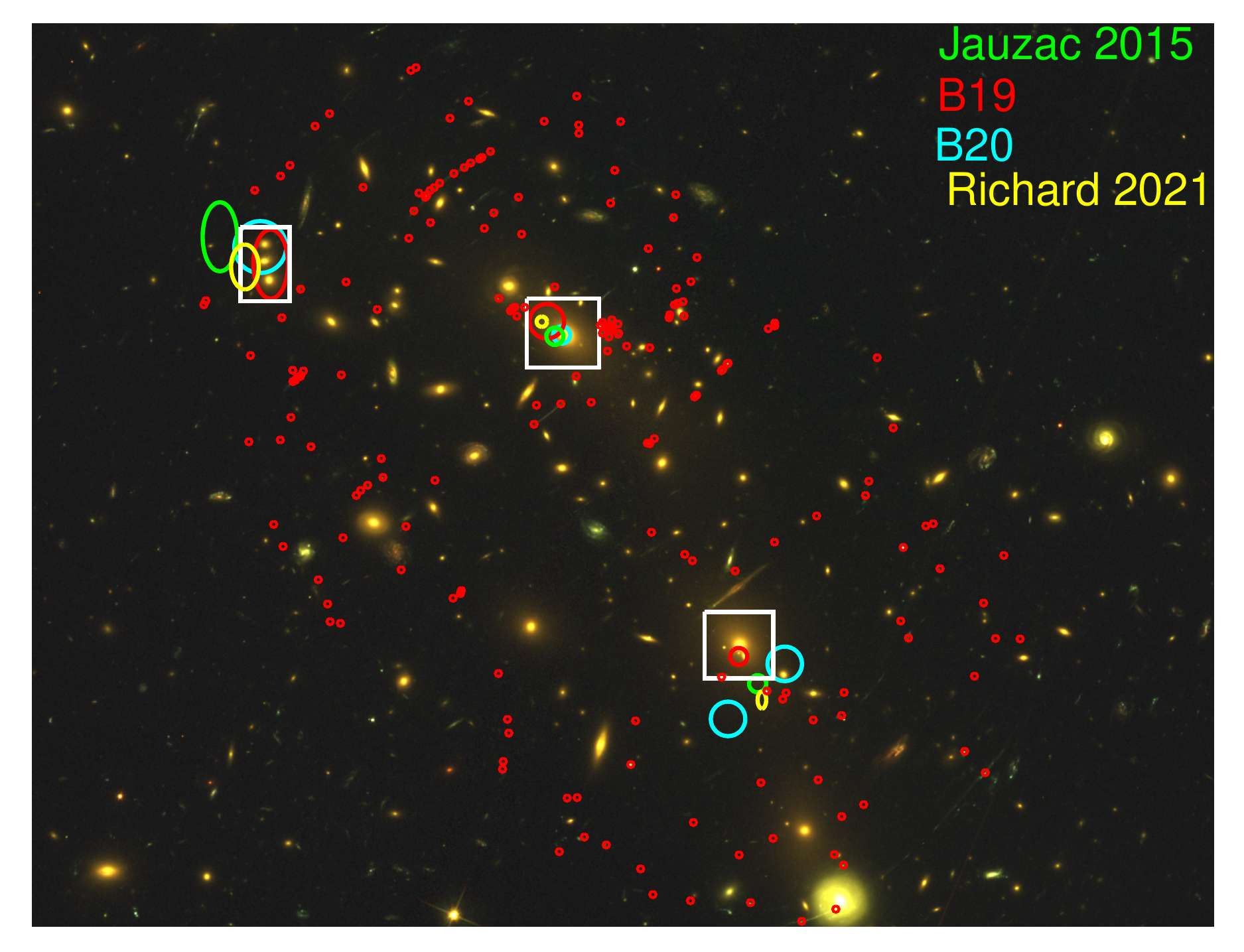}
\caption{Core of MACS\,0416. Red circles represent the multiple images used in this work (see B19 for details). Location of the mass clumps from different studies are shown by ellipses, 
whose size represent the 1$\sigma$ error bar on the position.
White boxes represent the priors on the position of each DM clump.
North is up, east is left. Size of the field equals to 137$\arcsec$\,$\times$\,105$\arcsec$.}
\label{0416Clumps}
\end{center}
\end{figure*}

B19 reproduces 107 multiple images using a 3 DM mass clumps model with 193 galaxy members 
in the modelling, reaching an RMS of 0.61$\arcsec$.
Each mass clump is associated with one of the three light concentrations of the cluster core
(Fig.~\ref{0416Clumps}): one in the centre, associated with the BCG;
one in the south west where a bright galaxy dominates;
and one in the north east where three cluster members are located.
They also optimize individually some parameters of two galaxies which are found
to have an influence on some multiple images. 
We keep them in the models explored here and refer the reader to B19 for details.

B21 reproduces 182 multiple images using a 4 DM mass clumps model with 212 galaxy members
in the modelling, reaching an RMS of 0.40$\arcsec$.
This fourth mass component is located in the south west of the cluster 
core (Fig.~\ref{0416Clumps}). This makes two 
mass clumps in this area, none of them 
being clearly associated with the brightest galaxy located in this area.
As discussed above, we consider that having DM clumps not associated with any
luminous counterpart is not satisfactory, especially in this case where it improves the fit
by only $\sim$0.1$\arcsec$.

Indeed, B21 also present three mass models which reproduce the 182 multiple images fairly
well (RMS between 0.45$\arcsec$ and 0.48$\arcsec$ instead of 0.40$\arcsec$) where the
DM is described using three mass clumps only.
Using three other figures of merit (matching of the internal kinematics of the cluster
members galaxies as well as BIC and AIC criteria), they choose the four mass clumps model as their
reference model.

We also reproduce such a mass model where the 182 images are reproduced by a three
mass clumps model with an RMS equal to 0.49$\arcsec$.
The location of each mass clump is allowed to vary by $\pm$ 15$\arcsec$ from its luminous
counterpart.
We find each mass clump to coincide with its luminous counterpart even if not imposed.
However, we note that the ellipticity of the two main DM clumps (the north east one being circular)
is above 0.7 (0.84 and 0.73 respectively).

We then impose the ellipticity of each mass clump to be smaller than 0.7, and their positions 
to coincide with the light distribution (boxes on Fig.~\ref{0416Clumps}).
For the central and south west mass clumps, positions
are allowed to vary within $\pm$ 4$\arcsec$ from the centre of the associated bright galaxy.
The north east mass clump position is allowed to vary within $\pm$ 3$\arcsec$ along the $x$ axis 
and within $\pm$ 4$\arcsec$ along the $y$ axis from the centre of the associated luminous counterpart.

The RMS equals to 0.63$\arcsec$.
The core radii of each mass clump is equal to 41.5$\pm$2.7; 51.2$\pm$3.7 and 60.4$\pm$8.6 kpc
respectively.
The ellipticity of the main and north-east mass distributions are stuck to 0.7.
This model constitutes the reference model from which we continue our investigations.
Its parameters are given in Table~\ref{0416table1}.

Similar three clumps mass models have also been proposed by \citet{0416hff} and \citet{Richard_2021}.
We show on Fig.~\ref{0416Clumps} the positions of the mass clumps inferred in their
analysis.

\begin{table*}
\begin{center}
\begin{tabular}{ccccccc}
\hline \\*[-1mm]
Cored Model & $\Delta$\,\textsc{ra} & $\Delta$\,\textsc{dec} &  $e$  & $\theta$ & $\sigma$  & $r_{\rm core}$ \\
            &  $\arcsec$            &  $\arcsec$             &       &          &  km\,s$^{-1}$ & kpc \\
\hline \\*[-1mm]
Clump 1  &  -1.8$\pm$0.5  & 0.4$\pm$0.3  & 0.69$\pm$0.01  & 143.3$\pm$1.0  &  629$\pm$16  & 41.5$\pm$2.7   \\   
\hline \\*[-1mm]
Clump 2  &  22.3$\pm$0.4  & -39.4$\pm$0.4  & 0.69$\pm$0.01  &  127.3$\pm$0.8  & 769$\pm$20 &  51.2$\pm$3.7  \\ 
\hline \\*[-1mm]
Clump 3  & -32.2$\pm$0.5  & 12$^{*}$  & 0.40$\pm$0.09  & 72$\pm$9 & 369$\pm$22  & 60.4$\pm$8.6  \\
\hline \\*[-1mm]
\end{tabular}
\end{center}
\caption{dPIE parameters inferred for the reference cored model of MACS\,0416, with an RMS equal to 0.63$\arcsec$.
Coordinates are given in arcseconds relative to $\alpha$\,=\,64.0381417, $\delta$\,=\,-24.0674722;
$e$ and $\theta$ are the ellipticity and position angle of the mass distribution.
Error bars correspond to the 1$\sigma$ confidence level.
Parameters with $*$ are stuck to a bound of the allowed prior.
For an L* galaxy, we have $\sigma$\,=\,237$\pm$10 km\,s$^{-1}$ and $r_s$\,=\,10$\pm$2\,{\rm{kpc}}.}
\label{0416table1}
\end{table*}

\subsection{Non cored mass model}
We redo the modelling imposing the core radius of each mass clump to be smaller than 10\,kpc.

We get an RMS equal to 2.07$\arcsec$.
This is 1.44$\arcsec$ worse than the cored mass model,
which is substantial. A cored mass model is therefore clearly preferred
but this difference of RMS is not enough to 
disentangle between a cored and a non cored model according to the 
(somehow arbitrary) criteria proposed in Section~2.4.
The ellipticity of the main and north-east mass distributions are also stuck to 0.7.
The core radii of the three mass clumps are stuck to 10\,kpc.
All parameters are given in Table~\ref{0416table_noncored}.

\begin{table*}
\begin{center}
\begin{tabular}{ccccccc}
\hline \\*[-1mm]
Non Cored Model & $\Delta$\,\textsc{ra} & $\Delta$\,\textsc{dec} &  $e$  & $\theta$ & $\sigma$  & $r_{\rm core}$ \\
            &  $\arcsec$            &  $\arcsec$             &       &          &  km\,s$^{-1}$ & kpc \\
\hline \\*[-1mm]
Clump 1  &  -1.1$\pm$0.3  & -0.1$\pm$0.2  & 0.7$^{*}$  & 135.4$\pm$0.6  &  522$\pm$4  & 10$^{*}$   \\
\hline \\*[-1mm]
Clump 2  &  21.6$\pm$0.2  & -37.4$\pm$0.4  &  0.7$^{*}$  &  129.6$\pm$0.8  & 651$\pm$5 &  10$^{*}$  \\
\hline \\*[-1mm]
Clump 3  & -32.1$\pm$0.1  & 12.9$\pm$0.1  & 0.27$\pm$0.08  & 126.5$\pm$9.0 & 357$\pm$8  & 10$^{*}$  \\
\hline \\*[-1mm]
\smallskip
\end{tabular}

\end{center}
\caption{Same as Table~\ref{0416table1} for the non cored model of MACS\,0416, with an RMS equal to 1.88$\arcsec$.
%Coordinates are given in arcseconds relative to $\alpha$\,=\,64.0381417, $\delta$\,=\,-24.0674722;
%$e$ and $\theta$ are the ellipticity and position angle of the mass distribution.
%Error bars correspond to the 1$\sigma$ confidence level.
%Parameters with $*$ are stuck to a bound of the allowed prior.
%Velocity dispersions ($\sigma$) are expressed in km/s and radii in kpc.
For an L* galaxy, we have $\sigma$\,=\,240$\pm$5 km\,s$^{-1}$ and $r_s$=108\,kpc (stuck to the 
upper bound of the allowed prior).}
\label{0416table_noncored}
\end{table*}

\subsection{Cosmology}
We run a cored and a non cored mass model, letting $w_{\rm X}$ and $\Omega_{\rm M}$ free.
For each model, we compare the values of the optimized output cosmology with the values of the reference
cosmology mentioned above.

For the cored model, we get RMS\,=\,0.51$\arcsec$ (instead of 0.63$\arcsec$ for the reference model) and the 
cosmology is off ($\Omega_{\rm M}$\,=\,0.16$\pm$1.0, $w_{\rm X}\,<$1.3).
For the non cored model, we get RMS\,=\,1.74$\arcsec$, and the cosmology
is totally off, basically unconstrained.

This test does not gives further insights to disentangle between a cored and a non cored mass model.

\section{MACS\,J1206}

MACS\,J1206 is a unimodal cD dominated galaxy cluster located at redshift 0.439.
Although considered as dynamically relaxed (Sereno 2017), the smooth component (DM + hot gas)
shows a significant asymmetry.

\subsection{Revisiting Bergamini et~al. (2019) model}
B19 reproduce 82 multiple images using a mass distribution composed of three dark matter haloes,
three dPIE clumps to model the X-ray surface brightness \citep{Bonamigo_2018}, a strong external shear ($\gamma_{\rm ext}=0.12$)
of unknown origin and
258 haloes describing the cluster members.
This set of multiple images is well reproduced (RMS\,=\,0.46$\arcsec$).

The positions of the three DM clumps are illustrated in Fig.~\ref{1206fig} (cyan ellipses). 
If one of these mass clumps is coincident with the cD galaxy, the others 
are not associated
with any luminous counterpart. 
Their core radii equal to 37, 82 and 69\,kpc (central, eastern and western clumps respectively). 
B19 state that these DM clumps are necessary to reproduce the apparent
elongated asymmetry of the cluster.
A similar mass model, in terms of number and position of DM clumps, has been presented by \citet{Richard_2021}, reaching an equivalent RMS.

Such a three clumps mass model was already proposed in the study by 
\citet{Caminha_2017}, who noted that these
"multi mass components should not be associated with extra DM haloes, but rather
to extra asymmetries or high order multipoles that a single parametric profile
cannot account for".

If we assume that the mass distribution is described by a single dPIE mass clump 
associated with the cD galaxy 
plus the galaxy component and the X-ray gas
and perform the mass modelling,
we end up with an RMS of 2.24$\arcsec$. 
Looking at the mass clump parameters, we find that the ellipticity is stuck to the
higher bound allowed (0.7) and that the core radius is constrained to
74$\pm$3 kpc.

The RMS difference between the former two models, equal to 1.8$\arcsec$, 
favors a three mass clumps model.
We consider that such a three clumps model is not satisfactory and we
conclude that the SL constraints in MACS\,J1206 are
\emph{not} well reproduced by a single halo described parametrically.

With the goal to present a physically motivated parametric mass
model for MACS\,J1206, we pursue our investigations.

\subsection{A two clumps mass model}

We notice that there is a bright galaxy east of the BCG (yellow box on Fig.~\ref{1206fig}) with which we associate a DM clump.
We impose the centre of this mass component to be within $\pm$\,2$\arcsec$ from the centre of the associated 
luminous component, i.e. within the yellow box on Fig.~\ref{1206fig}.
We also allow for an external shear component.

We end up with an RMS equal to 1.45$\arcsec$.
The main DM clump has a core radius equal to $\sim$\,35\,kpc and has a high but
reasonable ellipticity ($\sim$\,0.6).
On the other hand, the second DM clump is extremely flat, with a
core radius reaching the higher bound of the prior equal to 
230\,kpc.

\subsection{A three clumps mass model}
North west of the BCG is another bright galaxy (yellow box on Fig.~\ref{1206fig}) with which we associate a DM clump, investigating a
three clumps mass model.
We impose the centre of this mass component to be within $\pm$\,2$\arcsec$ from the centre of the associated
luminous component, and allow for an external shear component.

The resulting RMS equals to 1.38$\arcsec$. Parameters of the main
DM clump are compatible with the one obtained in the case of the two clumps mass model, but 
the two other DM clumps display very flat mass
profiles, with core radii equal to 190 and 230 kpc for the east
and north west clumps respectively.

\subsection{A four clumps mass model}
\begin{figure*}
\begin{center}
\includegraphics[scale=1.0,angle=0.0]{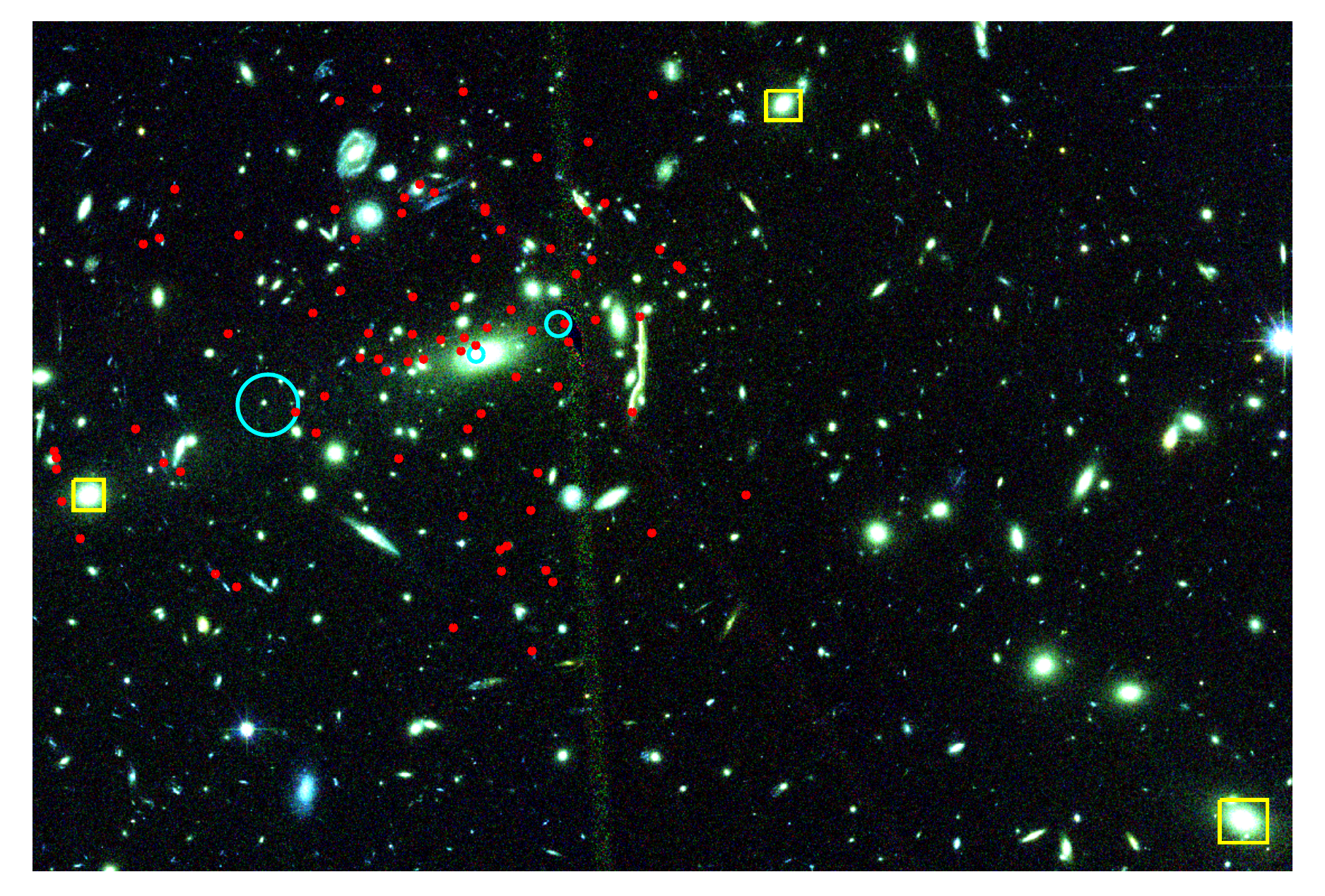}
\caption{Core of MACS\,J1206. Red circles represent the multiple images used in this work (see B19 for details). Location of the three mass clumps proposed by B19 are shown by cyan ellipses,
whose size represent the 1$\sigma$ error bar on the position.
Yellow boxes represent the prior on the position of the mass clumps included in the 
different mass model investigated.
North is up, east is left. Size of the field equals to 166$\arcsec$\,$\times$\,110$\arcsec$.}
\label{1206fig}
\end{center}
\end{figure*}

South west of the BCG is another bright galaxy (yellow box on Fig.~\ref{1206fig}) with which we associate a DM clump, investigating a
four clumps mass model.
We impose the centre of this mass component to be within $\pm$\,2$\arcsec$ from the centre of the associated
luminous component, and allow for an external shear component.

The resulting RMS equals to 1.19$\arcsec$.
Parameters of the main
DM clump are compatible with the one obtained in the case of the two clumps mass model.
The other three DM clumps display very flat mass profiles, with
core radii equal to 172, 200 and 145\,kpc respectively.

In all cases, we find that the inclusion of an external shear allows to lower the RMS by 
$\sim$\,0.6$\arcsec$. Its value is $\sim$\,0.06.

We see that we are able to lower the RMS by adding extra clumps which we associate with some
light concentrations. The more associations, the lower the RMS.
However, we are not able to get an RMS lower than 1$\arcsec$.
We also find that these extra mass clumps do require very large core
radii, between 145 and 230\,kpc.

We conclude from this serie of tests that MACS\,J1206 \emph{cannot} be reliably described by a pure
parametric model where each mass component would be associated with a luminous counterpart.

\section{Improving Parametric Mass Modelling with \emph{mild} Perturbations}
%bsplines

We see that in the case of MACS\,J1206, the parametric
approach reaches its limit and does not allow to provide a physically
motivated parametric description of the mass distribution.

We aim to see if adding a (mild) perturbation to the parametric modelling could
help to provide a decent fit.
Our main concern is whether adding this perturbation might make us lose
the advantages of the parametric mass modelling, i.e. if we end up with
clumps parameters that are biased by the perturbation, preventing us from
doing cluster physics.

Recently, something along this line has been implemented in \textsc{Lenstool}
\citep{Beauchesne_2021}.
This functionality consists in a surface of 2D B-spline functions added on top of the lensing potential.
While it has been shown to provide enhancement to the reconstruction of a simulated cluster, this is
the first time that it is applied on observational data.
In comparison to the free-form approach developed in \citet{jullo09}, this perturbative patch
cannot reproduce a full mass distribution and has to be combined with other analytical potentials. Indeed,
its total mass is always null, and it rather redistributes the one from the other model components. 
If complex grids of B-spline functions are possible, the current implementation is limited to a squared regular mesh. But
as the perturbation should be small in comparison to the total mass model, the induced biases can be 
neglected. Besides, the grid does not need more knowledge than the size of the constrained area to be built
such as an existing mass model or the light distribution.

We first test the inclusion of such perturbations on AS\,1063 which is already
well described by a parametric mass model, then we use them on MACS\,J1206.

\subsection{Test on AS\,1063}

We choose AS\,1063 in order to test this approach and see how the reference mass model
presented in Section~3.1 responds to the inclusion of the perturbation.

We perform a mass model for AS\,1063 as in Section~3, including the B-spline perturbation.
We use the same scheme of priors as in \citet{Beauchesne_2021} with a minimal lattice size of ~$105\arcsec$
and a grid of $5\times5$ basis functions. However, we kept the previous uniform priors for the dPIE parameters.
The resulting RMS equals to 0.36$\arcsec$, an improvement by almost a factor of two
when compared with the pure parametric mass model (0.67$\arcsec$).

We compare the parameters of the reference model (Table~1) with the parameters obtained when
adding the perturbation.
They do agree within the 3$\sigma$ error bars.
This test suggests that the B-spline perturbation is able to improve the pure parametric fit
\citep[as shown in][]{Beauchesne_2021}. 
Besides, this test also suggests that it is mild enough so that the parameters of the
associated parametric mass model are not significantly modified.

We also run a non cored mass model, including the B-spline perturbation.
The goal is to see if the perturbation is able to provide a decent non cored mass model.
In other words, is the perturbation able to compensate for the apparent need of a core in 
the centre?

\begin{figure*}
\begin{center}
\includegraphics[scale=0.9,angle=0.0]{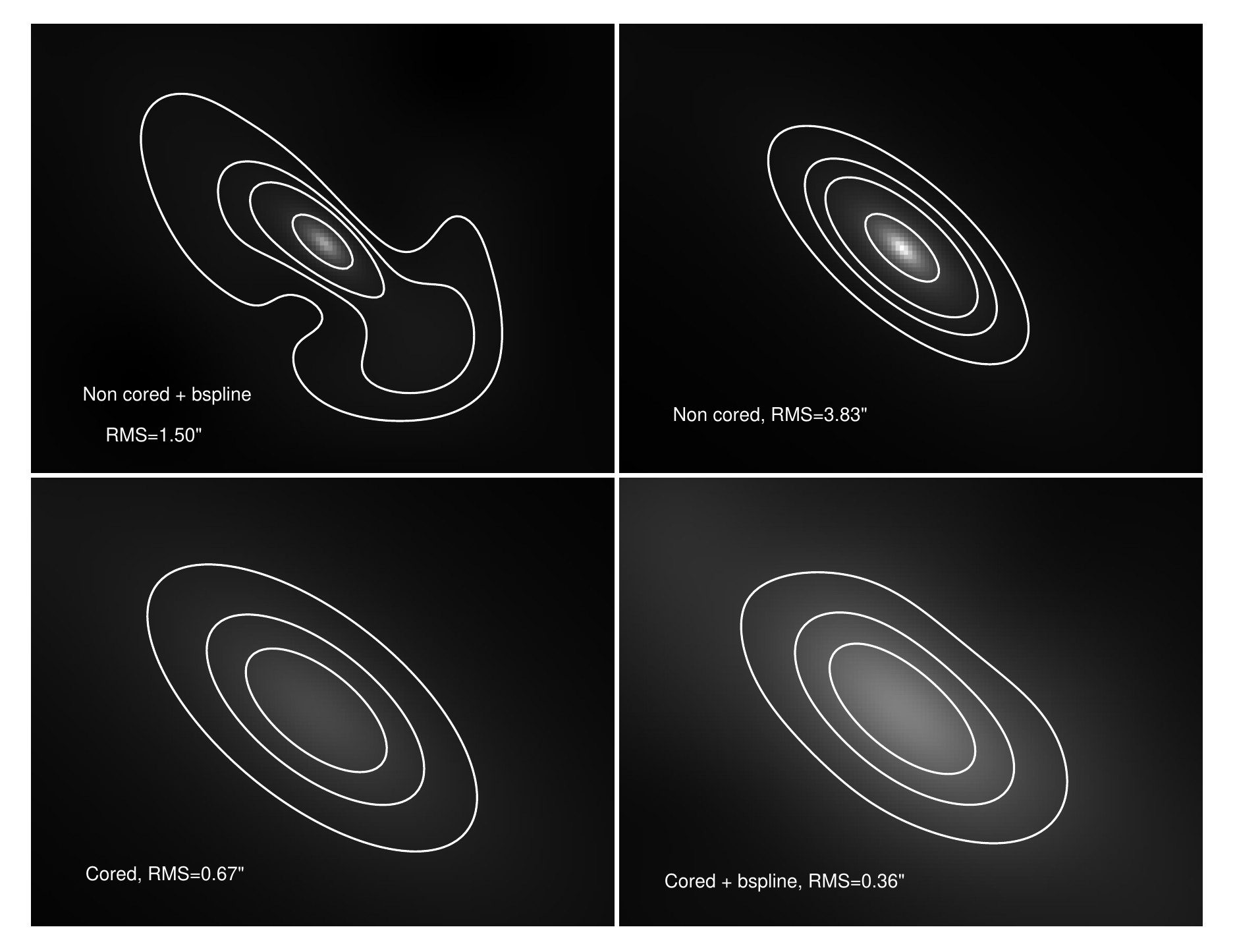}
\includegraphics[scale=0.9,angle=0.0]{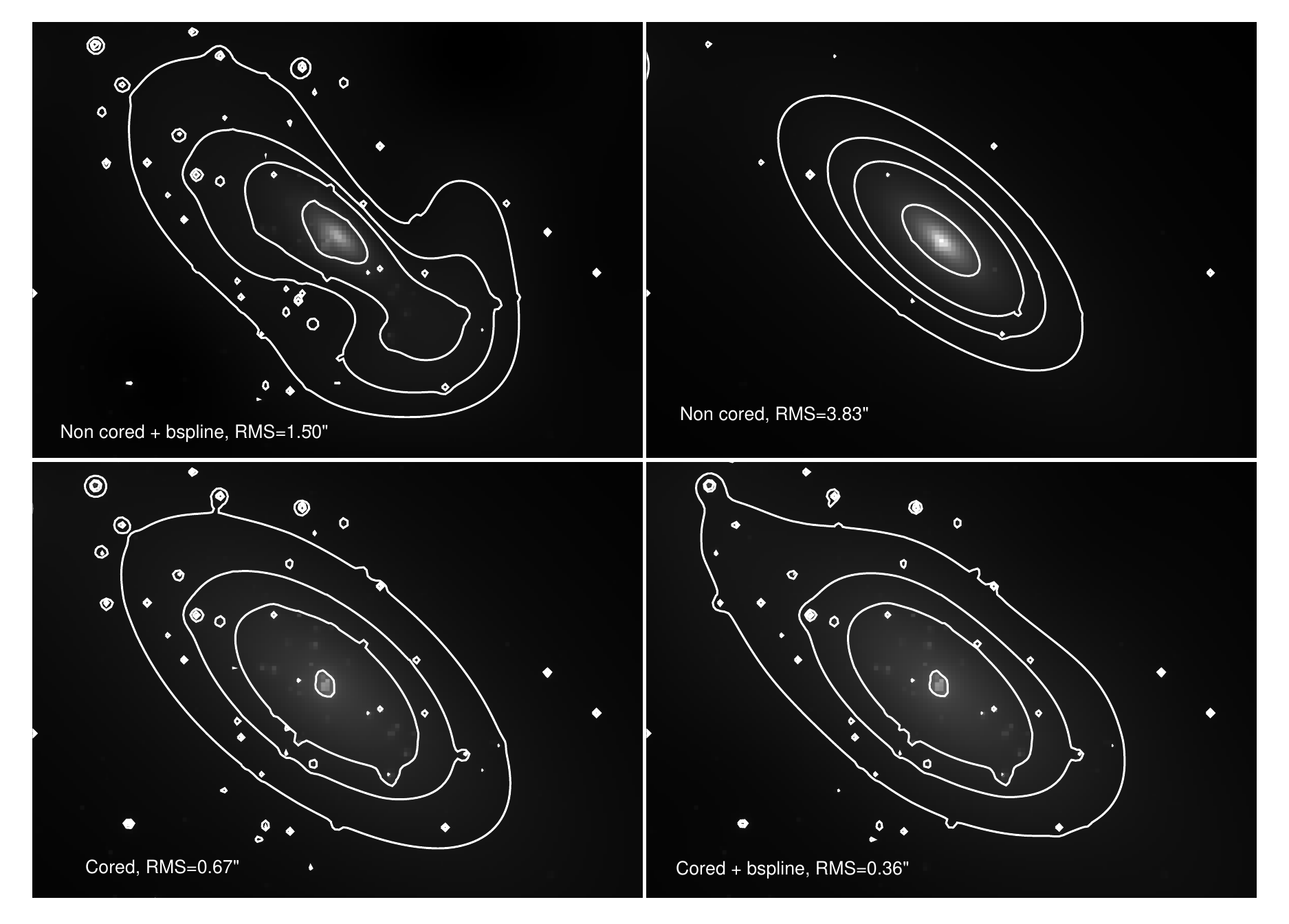}
\caption{2D mass maps (contours equal to 0.03, 0.04, 0.05 and 0.1 10$^{12}$ M$_{\sun}$\,arcsec$^{-2}$) for the different models explored here for AS\,1063. 
The size of each panel equals to 140$\arcsec$$\times$107$\arcsec$, centred on the cD galaxy.
\emph{Top:} we show the 2D mass distribution corresponding to the main DM halo + the B-spline
perturbation when included. 
\emph{Bottom:} on top of the main DM halo and the perturbation, we add the individual
galaxies.}
\label{1063_compare}
\end{center}
\end{figure*}

We find an RMS of 1.50$\arcsec$, instead of 3.83$\arcsec$ when not using the perturbation.
Comparing the outputs of these two models, we find that the main DM clump parameters
do agree with each other within the 3$\sigma$ error bars, once again suggesting that the
perturbation is mild enough not to significantly modify the parameters of the parametric
model.
Besides, given the value of the RMS, we find that the perturbation is not able to compensate for the need of a core in the centre.
In addition, the associated 2D mass map (Fig.~\ref{1063_compare}, upper left panel) 
has an irregular shape. If the parameters of the associated parametric model are not significantly
modified by the inclusion of the perturbation, we note that the resulting total mass map
appears unphysical.
Given the resulting shape of the mass map, we do not consider this mass model as "physically relevant"
and do not use it for further investigations.

We show the 2D mass maps associated with each model on Fig.~\ref{1063_compare}.
Considering the mass maps corresponding to the models including the perturbation, we see that when
the fit is good (RMS=0.36$\arcsec$), the shape of the mass contours are physically relevant, whereas
they look unphysical when the fit is bad (RMS=1.50$\arcsec$).

We also test this approach on MACS\,0416, investigating how the reference mass model
presented in Section~4.1 responds to the inclusion of the perturbation.
The resulting RMS equals to 0.46$\arcsec$ (versus 0.63$\arcsec$ for the reference model).
Besides, as for AS\,1063, the parameters obtained agree within the 3$\sigma$ error bars 
with the one obtained without the B-spline perturbation.

The results from these tests are encouraging enough to apply the perturbation on MACS\,J1206.

\subsection{Application on MACS\,J1206}

\subsubsection{A cored mass model allowing for the perturbation}

On top of the single central DM clump, we consider a B-spline perturbation.
No external shear is added.
We end up with an RMS equal to 0.53$\arcsec$ 
(instead of 2.24$\arcsec$ without
the perturbation nor external shear, see Section~5.1).

Allowing for an external shear component does not improve the fit.
B-spline coefficients differ, suggesting that an external shear component can be
described by the B-spline perturbation instead.
Comparing the parameters of the main DM clump obtained with and without the perturbation, 
we find that they do agree.
Indeed, 
comparing the output PDFs for each parameters, we find that all parameters but one 
do agree within the 3$\sigma$ error bars.
Only the position of the main clump along the X axis disagree between the two models.
This disagreement is small, equal to 5\,kpc.
We find that the perturbation is able to provide a decent fit without modifying the
parametric part, allowing to keep the advantages of the pure parametric mass modelling. 

Parameters of the main DM clump (when the perturbation is included) 
are given in Table~\ref{1206table}.
In particular, we do find a core radius equal to 
57.4$\pm$1.7\,kpc.

\subsubsection{Non cored mass model} 

Keeping the perturbation in the modelling, we then impose the core radius to be smaller 
than 10\,kpc.
The resulting RMS equals 7.39$\arcsec$.
Parameters are given in Table~\ref{1206table}.
The difference in RMS is substantial, equal to 6.83$\arcsec$, which is large enough to favor
a cored mass model for MACS\,J1206.
The associated mass map displays an unphysical shape (Fig.~\ref{1206_compare}).

\begin{table*}
\begin{center}
\begin{tabular}{cccccccccc}
\hline \\*[-1mm]
Model & $\Delta$\,\textsc{ra} & $\Delta$\,\textsc{dec} &  $e$  & $\theta$ & $\sigma$  & $r_{\rm core}$ & L$^{*}$ galaxy $\sigma$ & L$^{*}$ galaxy $r_s$ & RMS \\
&      $\arcsec$        &    $\arcsec$           &       &          & km\,s$^{-1}$ &  kpc     &  km\,s$^{-1}$             &      kpc               & $\arcsec$ \\
\hline \\*[-1mm]
Cored &  1.7$\pm$0.4  & 0.5$\pm$0.2 & 0.62$\pm$0.01  & 12.3$\pm$0.3  &  1071$\pm$7  & 57.4$\pm$1.7   &  266$\pm$6  & 41$\pm$4.6 &  0.53$\arcsec$ \\   
\hline \\*[-1mm]
Non Cored  &  1.7$\pm$0.2  & 1.3$\pm$0.2  &  0.7$^{*}$  &  18.2$\pm$1.0  & 909$\pm$8 &  10$^{*}$  &   253$\pm2$ & 92$\pm$12.0 & 7.39$\arcsec$  \\
\hline \\*[-1mm]
\smallskip
\end{tabular}
\end{center}
\caption{dPIE parameters inferred for the cored and the non cored models for MACS\,J1206, when the bspline perturbation is included.
Coordinates are given in arcseconds relative to $\alpha$\,=\,181.55062, $\delta$\,=\,-8.8009361;
$e$ and $\theta$ are the ellipticity and position angle of the mass distribution.
Error bars correspond to the 1$\sigma$ confidence level.
Parameters with $*$ are stuck to a bound of the allowed prior.}
\label{1206table}
\end{table*}

\subsubsection{Additional Tests}
Keeping the core radius smaller than 10\,kpc, we test what happens if we do not include the
B-spline perturbation.
We end up with an RMS equal to 11$\arcsec$.

Adding an external shear component lowers the RMS to 9.4$\arcsec$. The value of this external shear
is stuck to 0.1, the upper bound allowed by the prior (Section~2.3).
Interestingly, if we allow this external shear to reach higher values, it ends up to be constrained
to 0.26, a very unrealistic value since it is comparable to what would be experienced at 
100$\arcsec$ from the centre of galaxy cluster Abell~1689 \citep{mypaperIII}.
In this case the RMS lowers to 2.4$\arcsec$. This highlights once again the importance of providing
well motivated priors in SL mass modelling.

\begin{figure*}
\begin{center}
\includegraphics[scale=0.9,angle=0.0]{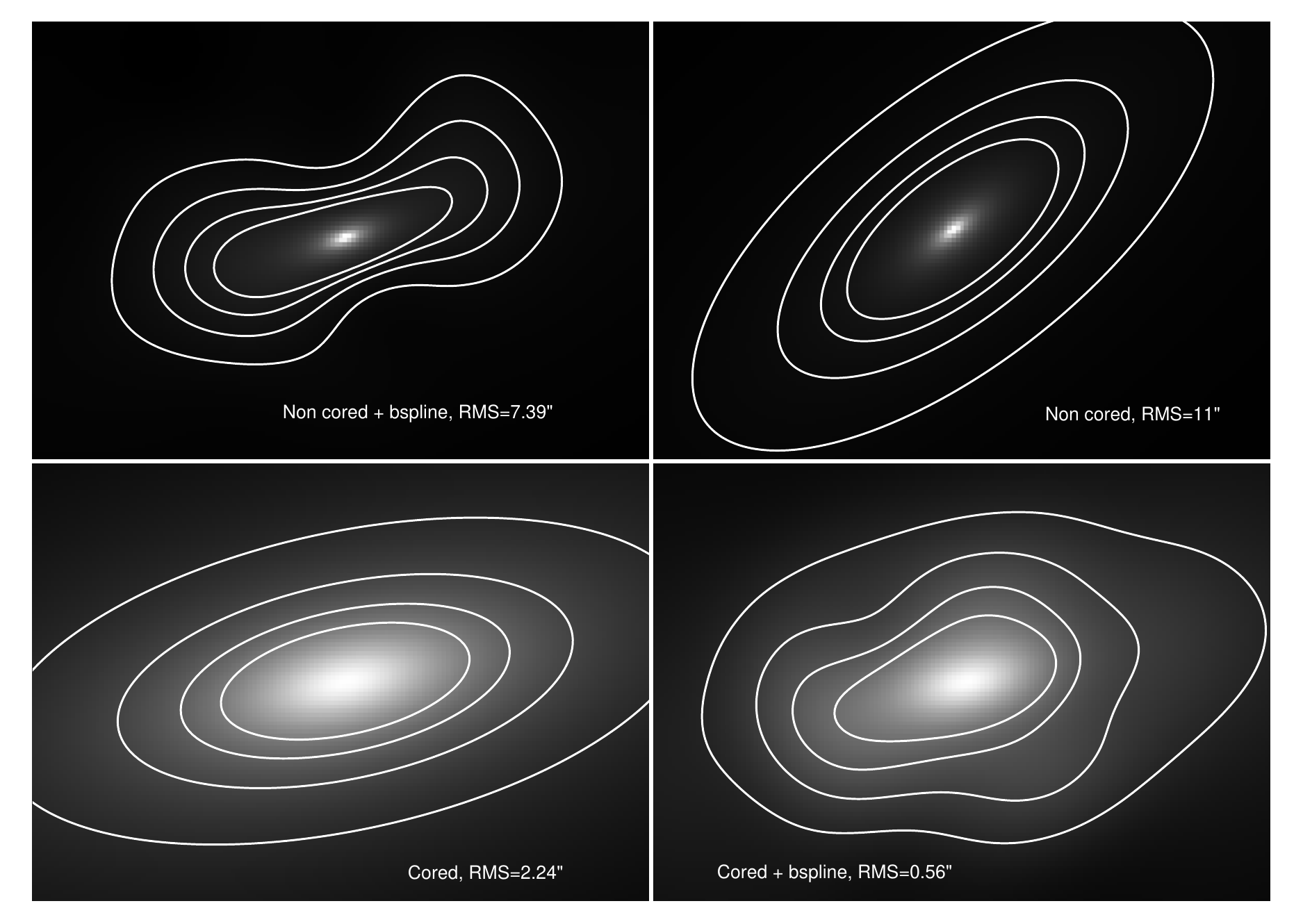}
\includegraphics[scale=0.9,angle=0.0]{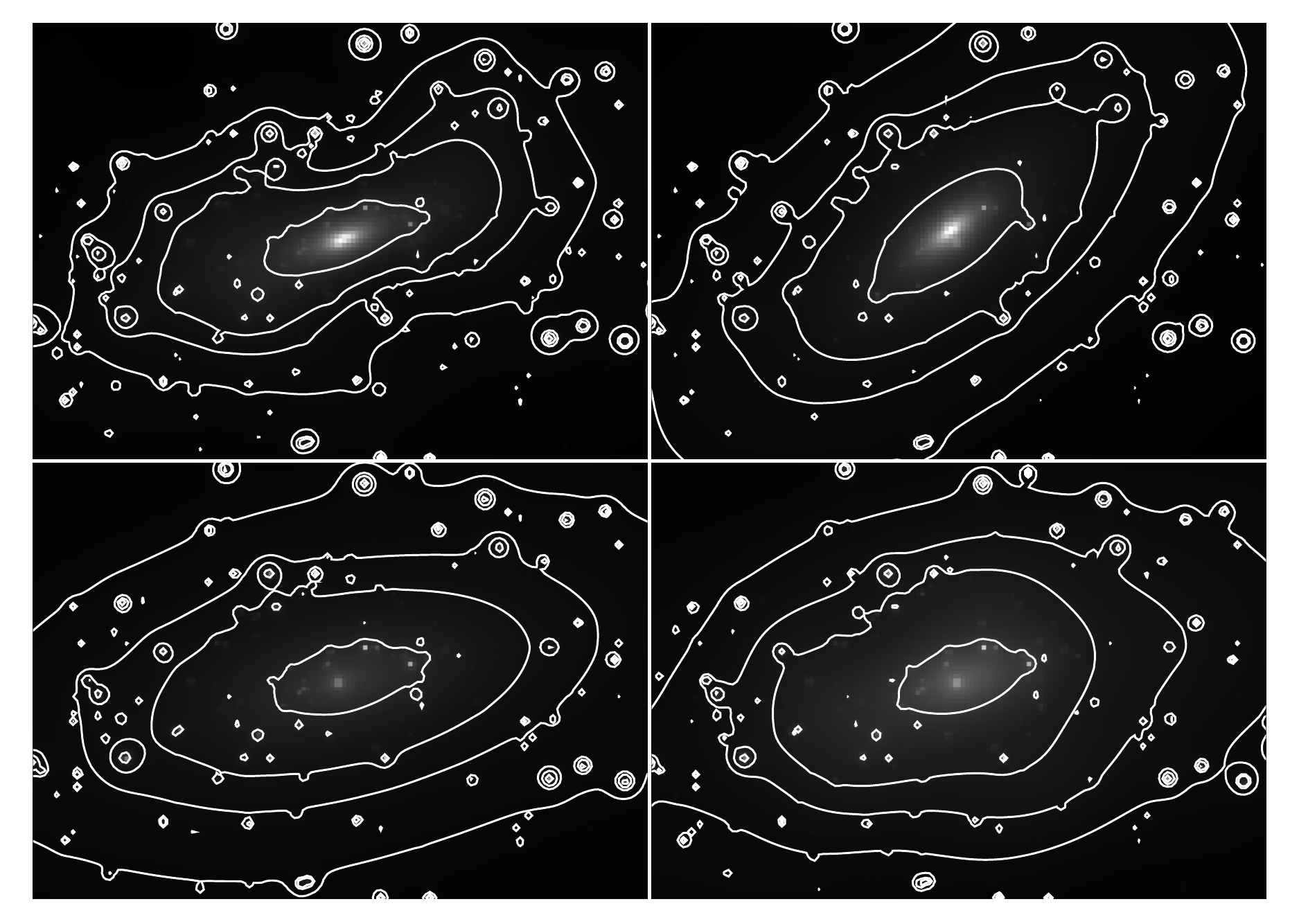}
\caption{2D mass maps for the different models explored here for MACS\,J1206.
For clarity, we label these models on the four top mass maps only. Same labels apply for the four bottom maps.
The size of each panel equals to
150$\arcsec$$\times$105$\arcsec$, centred on the cD galaxy.
\emph{Top:} we show the 2D mass distribution corresponding to the main DM halo + the B-spline
perturbation when included.
Contours equal to 0.02, 0.03, 0.04, and 0.05 10$^{12}$ M$_{\sun}$\,arcsec$^{-2}$.
\emph{Bottom:} on top of the main DM halo and the perturbation, we add the individual
galaxies.
Contours equal to 0.02, 0.03, 0.04, and 0.08 10$^{12}$ M$_{\sun}$\,arcsec$^{-2}$.}
\label{1206_compare}
\end{center}
\end{figure*}

In Fig.~\ref{1206_compare}, we show the 2D mass maps corresponding to the different models 
of MACS\,J1206 explored.
As for AS\,1063, considering the mass maps corresponding to the models including the perturbation, we see 
that when the fit is good (RMS=0.53$\arcsec$), the shape of the mass contours are physically relevant, 
whereas they look unphysical when the fit is bad (RMS=7.39$\arcsec$).

\subsection{Degeneracies with the BCG}
We remove the BCG from the galaxy catalog and optimise it explicitly, in order to investigate
any degeneracies between the main DM clump core radius and the BCG parameters.
The B-spline perturbation is considered in the modelling.
The position of the BCG is allowed to vary within $\pm$ 4$\arcsec$ from its centre and its core radius
between 1 and 50 kpc.
We use the values of B19 to put constraints on its velocity dispersion.
B19 provide a gaussian prior with a mean and a standard deviation. We consider the same
mean and the double of their standard deviation in order to allow for more freedom.

We get an RMS equal to 0.50$\arcsec$. The optimized position of the BCG is 
consistent with the light
distribution. Its ellipticity reaches the upper bound of the prior, and its core radius is constrained
to be smaller than 5\,kpc.
The parameters of the main DM halo are consistent with the one obtained without optimising the BCG
individually. Its core radius is found a bit larger in this case (14" instead of 13").
We see no degeneracies between the core radius of the BCG and the core radius of the main DM clump.

\subsection{Cosmology}

We consider the cored mass model presented in Table~\ref{1206table} and let the cosmology free.
The RMS equals to 0.53$\arcsec$ and the reference cosmology is retrieved
(Fig.~\ref{1206_compare_cosmo}, \emph{Left panel}).
Doing the same exercise with the non cored mass model, the RMS equals to 2.3$\arcsec$ (versus 7.39$\arcsec$).
The resulting cosmology is off, with $\Omega_{\rm M}$ $\sim$ 0.

This test gives additional support to the non cored mass model.

\subsection{Comparison with B19}

\begin{figure*}
\begin{center}
\includegraphics[scale=0.9,angle=0.0]{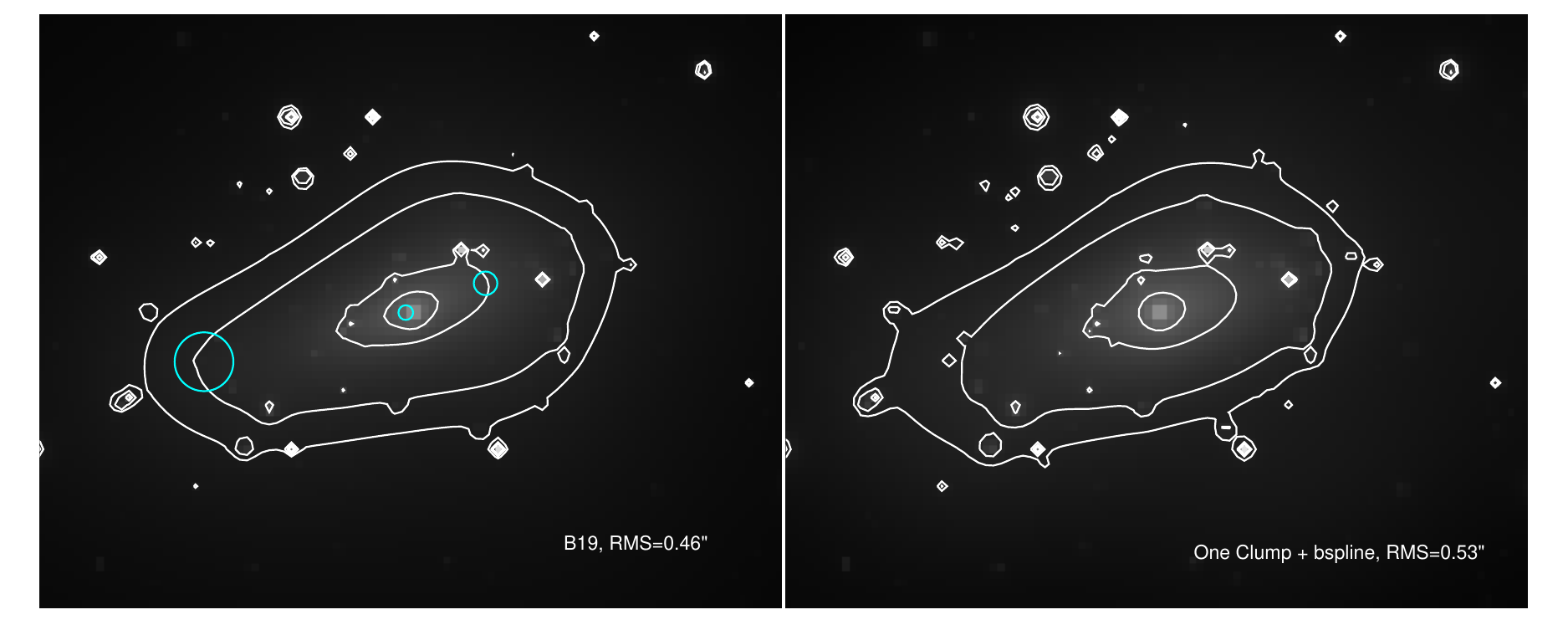}\\
\includegraphics[scale=0.5,angle=0.0]{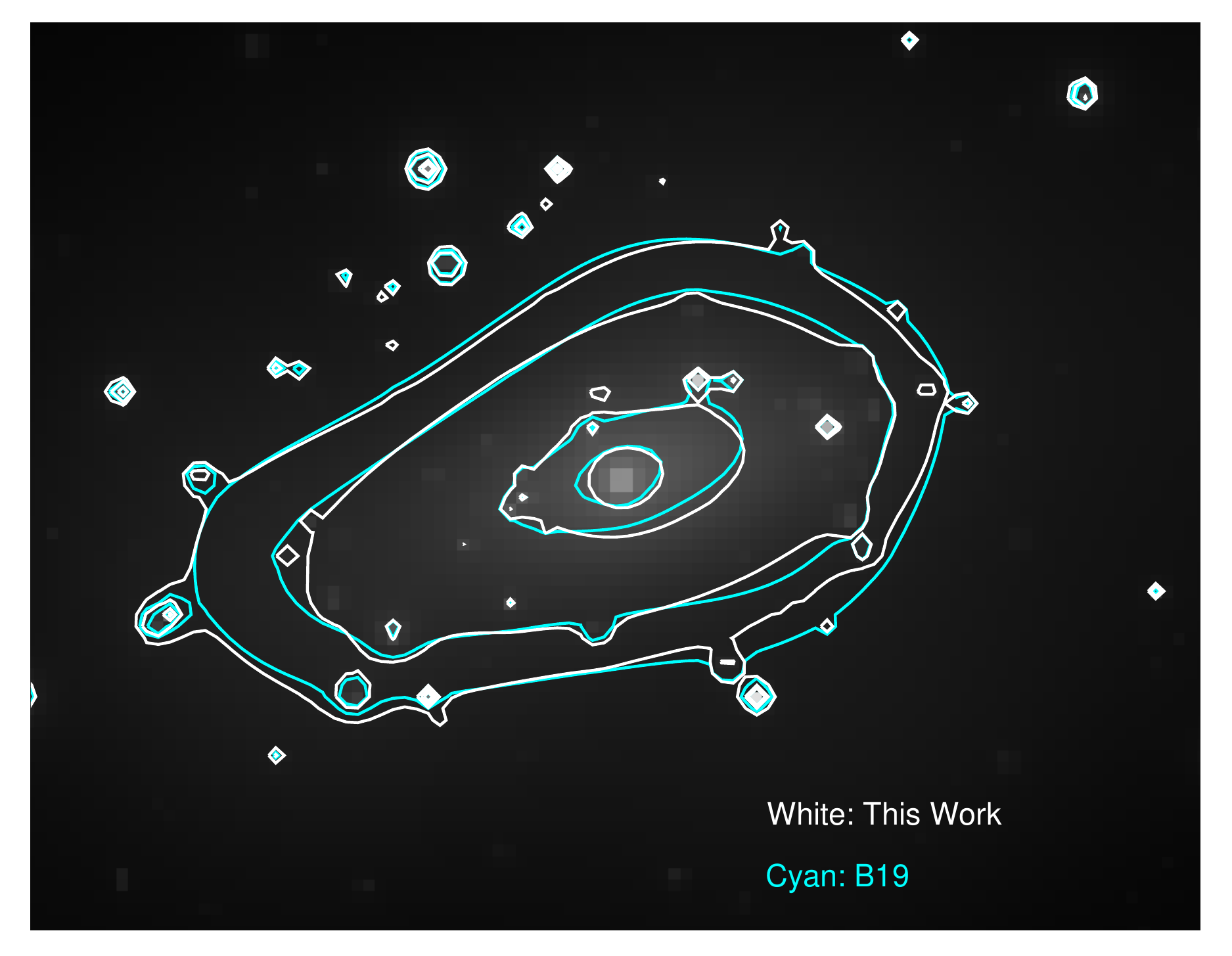}
\caption{Comparison between the 2D mass map presented in this work (\emph{Upper Right}) and the one obtained by B19 (\emph{Upper Left}) for MACS\,J1206. \emph{Lower Pannel:} We overplot the mass contours corresponding to both studies.
The size of each panel equals to
101$\arcsec$$\times$77$\arcsec$, centred on the cD galaxy.
Contours equal to 0.05, 0.06, 0.1, and 0.13 10$^{12}$ M$_{\sun}$\,arcsec$^{-2}$.
Cyan ellipses represent the location of the three mass clumps proposed by B19, as in Fig.~\ref{1206fig}.}
\label{M1206_compare}
\end{center}
\end{figure*}

In Fig.~\ref{M1206_compare}, we compare the 2D total mass map we obtain for MACS\,J1206
with the one proposed by B19.
We see that they have similar shapes: the model presented here is very close to the
one proposed by B19 in term of 2D total mass map.
This shows that the perturbation is able to reproduce the asymmetry given by the B19 three
clumps mass model using only one single mass clump.

\begin{figure}
\begin{center}
\includegraphics[scale=0.5,angle=0.0]{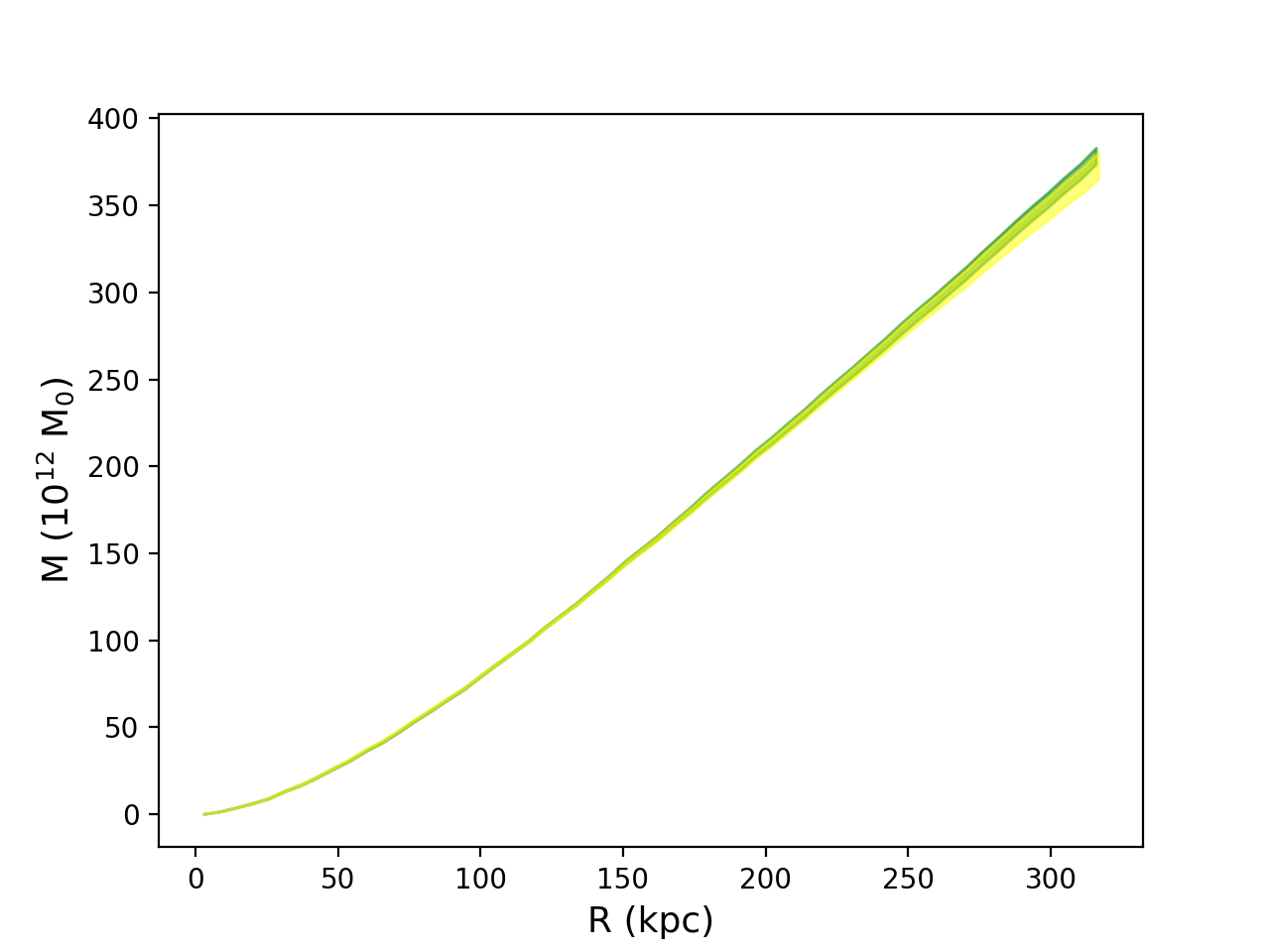}
\caption{Comparison between the integrated mass profile: B19 in green and the one  presented in this work in yellow. The 3$\sigma$ error bars are computed from the MCMC realizations.}
\label{Compar1206_Mass}
\end{center}
\end{figure}

We integrate the mass maps starting at the BCG centre in order to derive the 1D mass profile. 
The error bars are computed from the MCMC realizations.
We present the mass profile in the radial range where multiple images can be found, 
\emph{i.e.} up to $\sim$\,320 kpc from the centre.
As expected, we find they both agree with each other (Fig.~\ref{Compar1206_Mass}).

Both mass model fully agree in term of \emph{total mass} (both 2D maps and 1D integrated profile),
which is expected since the lensing constraints are sensitive to the total mass.
However, we have two very different ways of constructing this total mass, in term of the
underlying smooth DM component (the galaxy scale perturbers being fixed to the same values thanks to
the B19 results). 
When B19 propose a three clumps mass model plus an external shear component, 
we propose a one clump mass model with a perturbation.

\begin{figure*}
\begin{center}
\includegraphics[scale=0.9,angle=0.0]{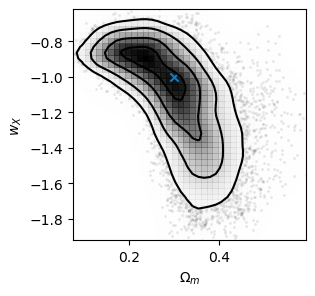}
\includegraphics[scale=0.9,angle=0.0]{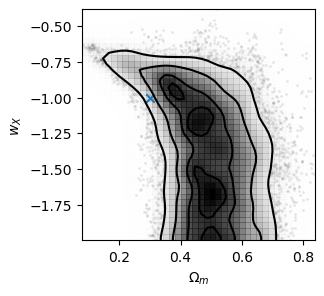}
\caption{Comparison between the constraints obtained on the cosmological parameters using the reference
model presented in Table~\ref{1206table} (\emph{Left}) and the one obtained when using the B19 model (\emph{Right}). The cyan cross represents the reference cosmology.}
\label{1206_compare_cosmo}
\end{center}
\end{figure*}

To further compare both models, we let the cosmological parameters free during the optimisation.
For the one clump cored mass model presented in this work, we consider the results obtained in Section~6.5 where
it is shown that the reference cosmology is retrieved.
We consider the mass model published by B19 and let the cosmology free during the optimisation.
To obtain stable results, we need to lower the \textsc{rate} value to 0.005.
We obtain an RMS equal to 0.43$\arcsec$, and the reference cosmology is not retrieved.
Constraints in the ($\Omega_{\rm M}, w_{\rm X}$) plane are compared in Fig.~\ref{1206_compare_cosmo}. 

These two mass models are likely to provide different estimates of the magnification.
Indeed, different mass distributions lead to different 
estimates of the magnification experienced by background sources. 
In MACS\,J0717, we found that the cored and the non cored mass models lead to different magnification estimates,
adding a systematic error that is in general larger than the statistical error derived from a given mass model.
A thorough investigation of the difference of magnification computed using the model presented here
and the one by B19 is beyond the scope of this paper, but it should be taken into account if using
MACS\,J1206 as a gravitational telescope, as it might decrease the area of the image plane where magnification
are well determined enough for reliable studies of the high-redshift Universe.

\section{Discussion}

We summarize the results presented in this paper in Table~\ref{resume}.
\begin{table*}
\begin{center}
\begin{tabular}{cccc}
\hline \\*[-1mm]
Model &  RMS & $\Delta(\rm RMS)$ & $r_{\rm core}$ \\
      &  $\arcsec$ & $\arcsec$ & kpc \\
\hline \\*[-1mm]
\hline \\*[-1mm]
AS\,1063, cored & 0.67 & 3.16 & 89.5$\pm$5.5 \\
\hline \\*[-1mm]
AS\,1063, non cored & 3.83 & --  & -- \\
\hline \\*[-1mm]
MACS\,0416, cored & 0.63 & 1.44  & 41.5$\pm$2.7; 51.2$\pm$3.7; 60.4$\pm$8.6 \\
\hline \\*[-1mm]
MACS\,0416, non cored & 1.88 & --  & -- \\
\hline \\*[-1mm]
MACS\,J1206, cored & 0.53 & 6.86  & 57.4$\pm$1.7 \\
\hline \\*[-1mm]
MACS\,J1206, non cored & 7.39 & --  & -- \\
\hline \\*[-1mm]
%MACS\,0717, cored & 1.9$\arcsec$ & 0.5$\arcsec$  & 104.3$\pm$13.0; 29.2$\pm$5.2; 178.8$\pm$13.6; 134.2$\pm$13.6 \\
%\hline \\*[-1mm]
%MACS\,0717, non cored & 2.4$\arcsec$ & --  & -- \\
%\hline \\*[-1mm]
\smallskip
\end{tabular}
\end{center}
\caption{Summary of the results presented in this paper.
For each cluster, we report the RMS obtained for both the cored and the
non cored model, as well as the difference of RMS between the non cored and the cored mass model.
B-spline perturbations are used in the case of MACS\,J1206 only.}
\label{resume}
\end{table*}

\subsection{Physically Motivated Parametric Mass Modelling}

In this work, we choose to present mass models which are physically motivated. 
In particular, we
want to avoid the inclusion of large scale DM clumps having no luminous counterpart, even if this might
degrade the SL fit.
Therefore we impose a strong prior on the position of the large scale DM clumps.
We prefer conclude that a parametric mass modelling is not adapted to describe a cluster instead
of using "adhoc dark clumps" whose physical interpretation is misleading. In these cases, a non parametric
approach might be considered instead.
In between, we have shown that using a mild perturbation in the form of B-splines is useful
to get a good fit while at the same time keeping the advantages of the 
pure parametric mass modelling.

Along this line, we have also considered other
well motivated priors, even though they do sometimes degrade the quality of the fit.
This is the case when including the X-ray component explicitly in AS\,1063.
This is also the case when limiting the strength of the external shear component in MACS\,J1206.

\subsection{On the inner shape and nature of DM}
We have investigated the inner shape of the DM distribution in a sample of
three massive galaxy clusters.
In \emph{all} cases, \emph{a cored mass distribution is preferred over a non cored
mass distribution.}
This is a potentially very exciting result, since it might be a signature of SIDM.
Prudently enough, we do not make any strong claims of evidence for SIDM.
This would require performing this kind of analyses on more galaxy clusters; doing rigorous 
tests with simulated data and exploring other ways to improve the models within the
standard Cold Dark Matter paradigm.
In particular, not fully understood interactions between baryons and dark matter can lead to 
core formation.

In two cases, the RMS difference is larger than 2$\arcsec$, which we consider
large enough in order to disentangle between a cored and a non cored mass distribution.
Actually it is even larger than 3$\arcsec$.
In AS\,1063, the RMS difference equals to 3.16$\arcsec$ and the core radius equals to 
89.5$\pm$5.5 kpc.
In MACS\,J1206, the RMS difference equals to 6.83$\arcsec$ and the core radius equals to
57.4$\pm$1.7 kpc. 
In MACS\,0416, the RMS difference equals to 1.44$\arcsec$ and the core radius of each clump equals to
41.5$\pm$2.7 kpc, 51.2$\pm$3.7 kpc and 60.4$\pm$8.6 kpc.

Note that we do not try to model AS\,1063 \& MACS\,J1206, the two unimodal clusters studied here,
using an NFW mass profile which is non cored by definition.
This is because we do not observe multiple images at radius large enough to constrain the typical 
size of the scale radius for a cluster scale NFW halo.
Ongoing BUFFALO observations \citep{buffalo}, in particular the weak lensing data, might help to test if a NFW mass
profile is able to reproduce the weak+strong lensing data for AS\,1063.

These values are summarized in Table~\ref{resume}. 
We note that some of them are at the higher bound (AS\,1063) 
of what seems to be 
allowed in an SIDM scenario.
As discussed in the introduction, a thorough investigation of core radii within an SIDM scenario is still
lacking.

Regarding SIDM, an observable consequence would be oscillations of the BCG around the centre
of the halo after mergers, which could persist for several Gyr \citep{Kim_2017}.
For the unimodal clusters studied here (AS\,1063 and MACS\,J1206), we find an offset between
the BCG and the DM smaller than $\sim$3\,kpc, which is compatible with a CDM scenario with no
self interactions \citep{Harvey_2019}.

Recent study by \citet{Andrade_2021} presented a SL analysis of AS\,1063 using a cored NFW mass
profile. It is not straightforward to compare the values of the core radius for such an NFW profile and 
the core radius of a dPIE profile.
However, \citet{Andrade_2021} do find evidence for a core radius equal to 19.83$^{+13.03}_{-9.41}$\,kpc.
These authors do study a sample of 8 regular galaxy clusters, out of which three do have a core
strictly larger than 0.

However, another recent study by \citet{Sartoris_2020} performing a joint fit to the velocity
dispersion profile of the BCG and to the velocity distribution of cluster member galaxies
over a radial range from 1\,kpc to the virial radius determine the inner slope of AS\,1063
to be fully consistent with the NFW predictions.

\subsection{Relaxing the B19 priors}

In MACS\,0717, it was found that the degeneracies between the smooth and the galaxy scale
components prevented to disentangle between a cored and a non cored mass model.
In this work we have been using priors on the galaxy scale component coming from spectroscopic
observations.
For each cluster studied here, we have relaxed the priors by B19, 
using flat priors for the velocity dispersion and scale
radius instead of the B19 results, as it is usually done in SL modelling.
The velocity dispersion is allowed to vary between 110 and 350 km\,s$^{-1}$
and the scale radius between 5 and 150\,kpc.

We present in Table~\ref{B19relax} the results for each cluster, for both the
cored and the non cored mass models, in terms of RMS and parameters of the galaxy scale
perturbers.

\begin{table*}
\begin{center}
\begin{tabular}{ccccc}
\hline \\*[-1mm]
Model &   $r_s$ & $\sigma$ & RMS & $\Delta(\rm RMS)$ \\
      &  kpc & km\,s$^{-1}$ & $\arcsec$ & $\arcsec$  \\
\hline \\*[-1mm]
\hline \\*[-1mm]
AS\,1063, cored & 150 (43)  & 230 (302)  & 0.64 (0.67) & 3.07 (3.16) \\
\hline \\*[-1mm]
AS\,1063, non cored & 5 (5) & 100 (303) & 3.71 (3.83) & - \\
\hline \\*[-1mm]
MACS\,0416, cored & 42 (10) & 264 (237) & 0.62 (0.63) & 1.18 (1.44) \\
\hline \\*[-1mm]
MACS\,0416, non cored & 42 (10) & 201 (237) & 1.80 (2.07) & - \\
\hline \\*[-1mm]
MACS\,J1206, cored & 29 (41) & 295 (266) & 0.52 (0.53) & 1.81 (6.83) \\
\hline \\*[-1mm]
MACS\,J1206, non cored & 150$^*$ (92)   & 297 (253) & 2.33 (7.49) & -\\
\hline \\*[-1mm]
%MACS\,0717, cored & 1.9$\arcsec$ & 0.5$\arcsec$  & 104.3$\pm$13.0; 29.2$\pm$5.2; 178.8$\pm$13.6; 134.2$\pm$13.6 \\
%\hline \\*[-1mm]
%MACS\,0717, non cored & 2.4$\arcsec$ & --  & -- \\
%\hline \\*[-1mm]
\smallskip
\end{tabular}
\end{center}
\caption{Results obtained on the galaxy scale perturbers when relaxing the B19 priors, for the
different models explored.
We also report the RMS obtained and the difference of RMS between the non cored and the cored mass model.
Values into brackets correponds to the results obtained when considering the B19 priors.}
\label{B19relax}
\end{table*}

In MACS\,J1206, we found that the B19 priors are important and actually allow to discriminate between a cored
and a non cored mass model.
We do not understand why these priors do help in MACS\,J1206 but not in AS\,1063 and MACS\,0416.
In the former two clusters, it is important to note that including these priors is 
essential since it provides more realistic parameters for the galaxy scale perturbers
and therefore gives more reliability to the results obtained on the inner shape of the DM
distribution in these clusters as we are confident that
they are not driven by the degeneracies between the smooth and the galaxy scale components.

We stress that this study by B19 is a major step forward and is paving the road towards
more realistic SL mass modelling.
With the development of spectrographs like MUSE, we will have more and more clusters
with spectroscopic data for galaxy scale perturbers to be included in SL modelling.

\subsection{Inclusion of the X-ray gas component}
For each cluster studied here, we do investigate a model where the X-ray gas component is not explicitly included in the
modelling.
For AS\,1063, the RMS equals 0.60$\arcsec$ instead of 0.67$\arcsec$ for the reference model.
The parameters of both models agree within 1$\sigma$, except for the velocity dispersion
of the main clump, which is about 100\,km\,s$^{-1}$ larger when the X-ray component is not
included.

For MACS\,0416, the RMS equals 0.57$\arcsec$ instead of 0.63$\arcsec$.
The parameters of both models for the different clumps agree within the 3$\sigma$ error bars.
The velocity dispersion of the north-east clump is found slightly higher
when the X-ray component is not included (437$\pm$34 vs 369$\pm$22 km/s).

For MACS\,1206, the RMS equals 0.67$\arcsec$ instead of 0.53$\arcsec$. 
The parameters of both models for the main clump agree within the 3$\sigma$ error bars, except for
the velocity dispersion, which is a bit larger to compensate (1158\,km\,s$^{-1}$ versus 1071\,km\,s$^{-1}$).

For AS\,1063 and MACS\,0416, the RMS is slightly better when not including the X-ray
gas component. 
Therefore, if looking at the best possible RMS, one might be tempted to favor a mass model 
where the X-ray component is not included.
However we observe this component in X-ray, therefore it is worth including it even if the
description is not the best possible.
This illustrates the question raised in the introduction of this paper about the balance between
having the best fit possible and presenting a physically motivated mass model.

Ultimately, we might want to fit simultaneously the lensing constraints and the X-ray data so that
the RMS becomes more representative of the inclusion of the X-ray component.

\subsection{On the choice of priors}

During an optimisation process, 
\textsc{Lenstool} always provides an answer, \emph{i.e.} some best fit parameters
describing a mass distribution. But \textsc{Lenstool}, as any other modelling algorithm,
does not care about the physical relevance of the outputs.
This is where expertise from the modeller comes into play.
Actually, we have seen that different modellers using the same set of data and the same
modelling algorithm can end up with different answers \citep{Meneghetti_2017}.

The parameter space being sometimes very large, it can be relevant to limit it
by imposing well motivated priors. Actually we always provide priors 
on each
parameters, \emph{i.e.} a range within which they are allowed to vary.
The tightness of this range depends on the availability of constraints coming from
other probes or from theory. 
For example, in this work we have used the constraints obtained by B19 in order
to put a prior on the galaxy scale perturbers.

We have seen at different occasions how the choice of priors does change the
best model inferred from the analysis.

Related to the choice of priors is the quest for the best fit. 
Of course we want the RMS to be as low as possible, and definitely below a given
threshold (assumed here to be 1$\arcsec$). However,
we have seen in this paper that sometimes a mass model can provide a better fit
but at the same time be unrealistic.

With this work, we aim to provide the following working hypothesis worth considering in
parametric mass modelling:\\
First, any large scale mass clump should be associated with a light counterpart.
Second, any well motivated prior should be included.

\section*{Acknowledgement}

We acknowledge P. Bergamini for sharing the results of the spectroscopic campaign which motivated us 
to revisit the mass models of these clusters.
We acknowledge the referee for a careful and constructive report.
ML acknowledges the Centre National de la Recherche Scientifique (CNRS) and the
Centre National des Etudes Spatiale (CNES) for support. 
This research has made use of computing facilities operated by CeSAM (Centre de donn\'ees Astrophysique de
Marseille) data centre at LAM, France (https://www.lam.fr/service/cesam/).

\bibliographystyle{aa} % style aa.bst
%\bibliography{master}
\bibliography{draft}

\begin{thebibliography}{50}
\expandafter\ifx\csname natexlab\endcsname\relax\def\natexlab#1{#1}\fi

\bibitem[{{Acebron} {et~al.}(2022){Acebron}, {Grillo}, {Bergamini}, {Mercurio},
  {Rosati}, {Caminha}, {Tozzi}, {Brammer}, {Meneghetti}, {Morelli}, {Nonino},
  \& {Vanzella}}]{Acebron_2022}
{Acebron}, A., {Grillo}, C., {Bergamini}, P., {et~al.} 2022, \apj, 926, 86

\bibitem[{Acebron {et~al.}(2017)Acebron, Jullo, Limousin, Tilquin, Giocoli,
  Jauzac, Mahler, \& Richard}]{Acebron_2017}
Acebron, A., Jullo, E., Limousin, M., {et~al.} 2017, Monthly Notices of the
  Royal Astronomical Society, 470, 1809–1825

\bibitem[{Acebron {et~al.}(2020)Acebron, Zitrin, Coe, Mahler, Sharon, Oguri,
  Bradač, Bradley, Frye, Forman, \& et~al.}]{Acebron_2020}
Acebron, A., Zitrin, A., Coe, D., {et~al.} 2020, The Astrophysical Journal,
  898, 6

\bibitem[{Andrade {et~al.}(2021)Andrade, Fuson, Gad-Nasr, Kong, Minor, Roberts,
  \& Kaplinghat}]{Andrade_2021}
Andrade, K.~E., Fuson, J., Gad-Nasr, S., {et~al.} 2021, Monthly Notices of the
  Royal Astronomical Society, 510, 54–81

\bibitem[{Beauchesne {et~al.}(2021)Beauchesne, Clément, Richard, \&
  Kneib}]{Beauchesne_2021}
Beauchesne, B., Clément, B., Richard, J., \& Kneib, J.-P. 2021, Monthly
  Notices of the Royal Astronomical Society, 506, 2002–2019

\bibitem[{Bergamini {et~al.}(2019)Bergamini, Rosati, Mercurio, Grillo, Caminha,
  Meneghetti, Agnello, Biviano, Calura, Giocoli, \& et~al.}]{Bergamini_2019}
Bergamini, P., Rosati, P., Mercurio, A., {et~al.} 2019, Astronomy \&
  Astrophysics, 631, A130

\bibitem[{Bergamini {et~al.}(2021)Bergamini, Rosati, Vanzella, Caminha, Grillo,
  Mercurio, Meneghetti, Angora, Calura, Nonino, \& et~al.}]{Bergamini_2021}
Bergamini, P., Rosati, P., Vanzella, E., {et~al.} 2021, Astronomy \&
  Astrophysics, 645, A140

\bibitem[{Bonamigo {et~al.}(2018)Bonamigo, Grillo, Ettori, Caminha, Rosati,
  Mercurio, Munari, Annunziatella, Balestra, \& Lombardi}]{Bonamigo_2018}
Bonamigo, M., Grillo, C., Ettori, S., {et~al.} 2018, The Astrophysical Journal,
  864, 98

\bibitem[{Caminha {et~al.}(2017)Caminha, Grillo, Rosati, Meneghetti, Mercurio,
  Ettori, Balestra, Biviano, Umetsu, Vanzella, \& et~al.}]{Caminha_2017}
Caminha, G.~B., Grillo, C., Rosati, P., {et~al.} 2017, Astronomy \&
  Astrophysics, 607, A93

\bibitem[{Caminha {et~al.}(2019)Caminha, Rosati, Grillo, Rosani, Caputi,
  Meneghetti, Mercurio, Balestra, Bergamini, Biviano, \& et~al.}]{Caminha_2019}
Caminha, G.~B., Rosati, P., Grillo, C., {et~al.} 2019, Astronomy \&
  Astrophysics, 632, A36

\bibitem[{Cerny {et~al.}(2018)Cerny, Sharon, Andrade-Santos, Avila, Bradač,
  Bradley, Carrasco, Coe, Czakon, Dawson, \& et~al.}]{RELICS_2018}
Cerny, C., Sharon, K., Andrade-Santos, F., {et~al.} 2018, The Astrophysical
  Journal, 859, 159

\bibitem[{Cibirka {et~al.}(2018)Cibirka, Acebron, Zitrin, Coe, Agulli,
  Andrade-Santos, Bradač, Frye, Livermore, Mahler, \& et~al.}]{Cibirka_2018}
Cibirka, N., Acebron, A., Zitrin, A., {et~al.} 2018, The Astrophysical Journal,
  863, 145

\bibitem[{Despali {et~al.}(2016)Despali, Giocoli, Bonamigo, Limousin, \&
  Tormen}]{Despali_2016}
Despali, G., Giocoli, C., Bonamigo, M., Limousin, M., \& Tormen, G. 2016,
  Monthly Notices of the Royal Astronomical Society, 466, 181–193

\bibitem[{{El{\'{\i}}asd{\'o}ttir} {et~al.}(2007){El{\'{\i}}asd{\'o}ttir},
  {Limousin}, {Richard}, {Hjorth}, {Kneib}, {Natarajan}, {Pedersen}, {Jullo},
  \& {Paraficz}}]{ardis2218}
{El{\'{\i}}asd{\'o}ttir}, {\'A}., {Limousin}, M., {Richard}, J., {et~al.} 2007,
  ArXiv e-prints, 710

\bibitem[{Fischer {et~al.}(2021)Fischer, Brüggen, Schmidt-Hoberg, Dolag,
  Kahlhoefer, Ragagnin, \& Robertson}]{Fischer_2021}
Fischer, M.~S., Brüggen, M., Schmidt-Hoberg, K., {et~al.} 2021, Monthly
  Notices of the Royal Astronomical Society, 505, 851–868

\bibitem[{Harvey {et~al.}(2019)Harvey, Robertson, Massey, \&
  McCarthy}]{Harvey_2019}
Harvey, D., Robertson, A., Massey, R., \& McCarthy, I.~G. 2019, Monthly Notices
  of the Royal Astronomical Society, 488, 1572–1579

\bibitem[{{Host}(2012)}]{host}
{Host}, O. 2012, \mnras, 420, L18

\bibitem[{Irastorza \& Redondo(2018)}]{Axions_review}
Irastorza, I.~G. \& Redondo, J. 2018, Progress in Particle and Nuclear Physics,
  102, 89–159

\bibitem[{{Jauzac} {et~al.}(2014){Jauzac}, {Cl{\'e}ment}, {Limousin},
  {Richard}, {Jullo}, {Ebeling}, {Atek}, {Kneib}, {Knowles}, {Natarajan},
  {Eckert}, {Egami}, {Massey}, \& {Rexroth}}]{0416hff}
{Jauzac}, M., {Cl{\'e}ment}, B., {Limousin}, M., {et~al.} 2014, \mnras, 443,
  1549

\bibitem[{{Jauzac} {et~al.}(2015){Jauzac}, {Richard}, {Jullo}, {Cl{\'e}ment},
  {Limousin}, {Kneib}, {Ebeling}, {Natarajan}, {Rodney}, {Atek}, {Massey},
  {Eckert}, {Egami}, \& {Rexroth}}]{jauzac2744}
{Jauzac}, M., {Richard}, J., {Jullo}, E., {et~al.} 2015, \mnras, 452, 1437

\bibitem[{{Jullo} \& {Kneib}(2009)}]{jullo09}
{Jullo}, E. \& {Kneib}, J. 2009, \mnras, 395, 1319

\bibitem[{{Jullo} {et~al.}(2007){Jullo}, {Kneib}, {Limousin},
  {El{\'{\i}}asd{\'o}ttir}, {Marshall}, \& {Verdugo}}]{jullo07}
{Jullo}, E., {Kneib}, J.-P., {Limousin}, M., {et~al.} 2007, New Journal of
  Physics, 9, 447

\bibitem[{Kim {et~al.}(2017)Kim, Peter, \& Wittman}]{Kim_2017}
Kim, S.~Y., Peter, A. H.~G., \& Wittman, D. 2017, Monthly Notices of the Royal
  Astronomical Society, 469, 1414–1444

\bibitem[{Lagattuta {et~al.}(2019)Lagattuta, Richard, Bauer, Clément, Mahler,
  Soucail, Carton, Kneib, Laporte, Martinez, \& et~al.}]{Lagattuta_2019}
Lagattuta, D.~J., Richard, J., Bauer, F.~E., {et~al.} 2019, Monthly Notices of
  the Royal Astronomical Society

\bibitem[{{Limousin} {et~al.}(2007{\natexlab{a}}){Limousin}, {Kneib},
  {Bardeau}, {Natarajan}, {Czoske}, {Smail}, {Ebeling}, \& {Smith}}]{mypaperII}
{Limousin}, M., {Kneib}, J.~P., {Bardeau}, S., {et~al.} 2007{\natexlab{a}},
  \aap, 461, 881

\bibitem[{{Limousin} {et~al.}(2005){Limousin}, {Kneib}, \&
  {Natarajan}}]{mypaperI}
{Limousin}, M., {Kneib}, J.-P., \& {Natarajan}, P. 2005, \mnras, 356, 309

\bibitem[{Limousin {et~al.}(2016)Limousin, Richard, Jullo, Jauzac, Ebeling,
  Bonamigo, Alavi, Clément, Giocoli, Kneib, \& et~al.}]{Limousin2016}
Limousin, M., Richard, J., Jullo, E., {et~al.} 2016, Astronomy \& Astrophysics,
  588, A99

\bibitem[{{Limousin} {et~al.}(2007{\natexlab{b}}){Limousin}, {Richard},
  {Jullo}, {Kneib}, {Fort}, {Soucail}, {El{\'{\i}}asd{\'o}ttir}, {Natarajan},
  {Ellis}, {Smail}, {Czoske}, {Smith}, {Hudelot}, {Bardeau}, {Ebeling},
  {Egami}, \& {Knudsen}}]{mypaperIII}
{Limousin}, M., {Richard}, J., {Jullo}, E., {et~al.} 2007{\natexlab{b}}, \apj,
  668, 643

\bibitem[{Lotz {et~al.}(2017)Lotz, Koekemoer, Coe, Grogin, Capak, Mack,
  Anderson, Avila, Barker, Borncamp, \& et~al.}]{Lotz_2017}
Lotz, J.~M., Koekemoer, A., Coe, D., {et~al.} 2017, The Astrophysical Journal,
  837, 97

\bibitem[{Mahler {et~al.}(2017)Mahler, Richard, Clément, Lagattuta, Schmidt,
  Patrício, Soucail, Bacon, Pello, Bouwens, \& et~al.}]{Mahler_2017}
Mahler, G., Richard, J., Clément, B., {et~al.} 2017, Monthly Notices of the
  Royal Astronomical Society, 473, 663–692

\bibitem[{Mahler {et~al.}(2019)Mahler, Sharon, Fox, Coe, Jauzac, Strait, Edge,
  Acebron, Andrade-Santos, Avila, \& et~al.}]{Mahler_2019}
Mahler, G., Sharon, K., Fox, C., {et~al.} 2019, The Astrophysical Journal, 873,
  96

\bibitem[{McCarthy {et~al.}(2016)McCarthy, Schaye, Bird, \&
  Le~Brun}]{McCarthy_2016}
McCarthy, I.~G., Schaye, J., Bird, S., \& Le~Brun, A. M.~C. 2016, Monthly
  Notices of the Royal Astronomical Society, 465, 2936–2965

\bibitem[{Meneghetti {et~al.}(2017)Meneghetti, Natarajan, Coe, Contini,
  De~Lucia, Giocoli, Acebron, Borgani, Bradac, Diego, \&
  et~al.}]{Meneghetti_2017}
Meneghetti, M., Natarajan, P., Coe, D., {et~al.} 2017, Monthly Notices of the
  Royal Astronomical Society, 472, 3177–3216

\bibitem[{Monna {et~al.}(2016)Monna, Seitz, Geller, Zitrin, Mercurio, Suyu,
  Postman, Fabricant, Hwang, \& Koekemoer}]{Monna_2016}
Monna, A., Seitz, S., Geller, M.~J., {et~al.} 2016, Monthly Notices of the
  Royal Astronomical Society, 465, 4589–4601

\bibitem[{{Natarajan} {et~al.}(2009){Natarajan}, {Kneib}, {Smail}, {Treu},
  {Ellis}, {Moran}, {Limousin}, \& {Czoske}}]{priya4}
{Natarajan}, P., {Kneib}, J.-P., {Smail}, I., {et~al.} 2009, \apj, 693, 970

\bibitem[{Newman {et~al.}(2013)Newman, Treu, Ellis, \& Sand}]{Newman_2013}
Newman, A.~B., Treu, T., Ellis, R.~S., \& Sand, D.~J. 2013, The Astrophysical
  Journal, 765, 25

\bibitem[{Paraficz {et~al.}(2016)Paraficz, Kneib, Richard, Morandi, Limousin,
  Jullo, \& Martinez}]{danka}
Paraficz, D., Kneib, J.-P., Richard, J., {et~al.} 2016, Astronomy \&
  Astrophysics, 594, A121

\bibitem[{{Postman} {et~al.}(2012){Postman}, {Coe}, {Ben{\'{\i}}tez},
  {Bradley}, {Broadhurst}, {Donahue}, {Ford}, {Graur}, {Graves}, {Jouvel},
  {Koekemoer}, {Lemze}, {Medezinski}, {Molino}, {Moustakas}, {Ogaz}, {Riess},
  {Rodney}, {Rosati}, {Umetsu}, {Zheng}, {Zitrin}, {Bartelmann}, {Bouwens},
  {Czakon}, {Golwala}, {Host}, {Infante}, {Jha}, {Jimenez-Teja}, {Kelson},
  {Lahav}, {Lazkoz}, {Maoz}, {McCully}, {Melchior}, {Meneghetti}, {Merten},
  {Moustakas}, {Nonino}, {Patel}, {Reg{\"o}s}, {Sayers}, {Seitz}, \& {Van der
  Wel}}]{clash}
{Postman}, M., {Coe}, D., {Ben{\'{\i}}tez}, N., {et~al.} 2012, \apjs, 199, 25

\bibitem[{Rahaman {et~al.}(2021)Rahaman, Raja, Datta, Burns, Alden, \&
  Rapetti}]{Rahaman_2021}
Rahaman, M., Raja, R., Datta, A., {et~al.} 2021, Monthly Notices of the Royal
  Astronomical Society, 505, 480–491

\bibitem[{Rescigno {et~al.}(2020)Rescigno, Grillo, Lombardi, Rosati, Caminha,
  Meneghetti, Mercurio, Bergamini, \& Coe}]{Rescigno_2020}
Rescigno, U., Grillo, C., Lombardi, M., {et~al.} 2020, Astronomy \&
  Astrophysics, 635, A98

\bibitem[{Richard {et~al.}(2021)Richard, Claeyssens, Lagattuta, Guaita, Bauer,
  Pello, Carton, Bacon, Soucail, Lyon, \& et~al.}]{Richard_2021}
Richard, J., Claeyssens, A., Lagattuta, D., {et~al.} 2021, Astronomy \&
  Astrophysics, 646, A83

\bibitem[{{Richard} {et~al.}(2010){Richard}, {Smith}, {Kneib}, {Ellis},
  {Sanderson}, {Pei}, {Targett}, {Sand}, {Swinbank}, {Dannerbauer}, {Mazzotta},
  {Limousin}, {Egami}, {Jullo}, {Hamilton-Morris}, \& {Moran}}]{locuss}
{Richard}, J., {Smith}, G.~P., {Kneib}, J., {et~al.} 2010, \mnras, 404, 325

\bibitem[{Robertson {et~al.}(2017)Robertson, Massey, \& Eke}]{Robertson_2017}
Robertson, A., Massey, R., \& Eke, V. 2017, Monthly Notices of the Royal
  Astronomical Society, 467, 4719–4730

\bibitem[{Rocha {et~al.}(2013)Rocha, Peter, Bullock, Kaplinghat,
  Garrison-Kimmel, Oñorbe, \& Moustakas}]{Rocha_2013}
Rocha, M., Peter, A. H.~G., Bullock, J.~S., {et~al.} 2013, Monthly Notices of
  the Royal Astronomical Society, 430, 81–104

\bibitem[{{Sand} {et~al.}(2004){Sand}, {Treu}, {Smith}, \& {Ellis}}]{sand04}
{Sand}, D.~J., {Treu}, T., {Smith}, G.~P., \& {Ellis}, R.~S. 2004, \apj, 604,
  88

\bibitem[{Sartoris {et~al.}(2020)Sartoris, Biviano, Rosati, Mercurio, Grillo,
  Ettori, Nonino, Umetsu, Bergamini, Caminha, \& et~al.}]{Sartoris_2020}
Sartoris, B., Biviano, A., Rosati, P., {et~al.} 2020, Astronomy \& Astrophysics,
  637, A34

\bibitem[{Schumann(2019)}]{direct_detection}
Schumann, M. 2019, Journal of Physics G: Nuclear and Particle Physics, 46,
  103003

\bibitem[{Steinhardt {et~al.}(2020)Steinhardt, Jauzac, Acebron, Atek, Capak,
  Davidzon, Eckert, Harvey, Koekemoer, Lagos, \& et~al.}]{buffalo}
Steinhardt, C.~L., Jauzac, M., Acebron, A., {et~al.} 2020, The Astrophysical
  Journal Supplement Series, 247, 64

\bibitem[{Tulin \& Yu(2018)}]{SIDM_review}
Tulin, S. \& Yu, H.-B. 2018, Physics Reports, 730, 1–57

\bibitem[{Zitrin(2021)}]{Zitrin_2021}
Zitrin, A. 2021, The Astrophysical Journal, 919, 54

\end{thebibliography}

\end{document}